\documentclass[apj,onecolumn]{emulateapj}
\usepackage{graphicx,natbib,color,amsmath,subfigure}
\usepackage{ulem}
\begin{document}
\title{Theory of gyroresonance and free-free emissions from non-Maxwellian quasi-steady-state electron distributions}
\author{Gregory D. Fleishman\altaffilmark{1,2,3}, Alexey A. Kuznetsov\altaffilmark{4}}
\altaffiltext{1}{Center For Solar-Terrestrial Research, New Jersey Institute of Technology, Newark, NJ 07102}
\altaffiltext{2}{Ioffe Physico-Technical Institute, St. Petersburg 194021, Russia}
\altaffiltext{3}{Central Astronomical Observatory at Pulkovo of RAS, Saint-Petersburg 196140, Russia}
\altaffiltext{4}{Institute of Solar-Terrestrial Physics, Irkutsk 664033, Russia}

\begin{abstract}

Currently there is a concern about ability of the classical thermal (Maxwellian) distribution to describe quasi-steady-state plasma in solar atmosphere including active regions. In particular, other distributions have been proposed to better fit observations, for example, kappa- and $n$-distributions. If present, these distributions will generate radio emissions with different observable properties compared with the classical gyroresonance (GR) or free-free emission, which implies a way of remote detecting these non-Maxwellian distributions in the radio observations. Here we present analytically derived GR and free-free emissivities and absorption coefficients for the kappa- and $n$-distributions and discuss their properties, which are in fact remarkably different from each other and from the classical Maxwellian plasma.  In particular, the radio brightness temperature from a gyrolayer increases with the optical depth $\tau$ for kappa-distribution, but decreases with $\tau$ for $n$-distribution. This property has a remarkable consequence allowing a straightforward observational test: the gyroresonance radio emission  from the non-Maxwellian distributions is supposed to be noticeably polarized even in the optically thick case, where the emission would have strictly zero polarization in the case of Maxwellian plasma. This offers a way of remote probing the plasma distribution in astrophysical sources including solar active regions as a vivid example.

\end{abstract}

%
\clearpage

\section{Introduction}

In recent years a critical mass of observationally-driven concerns about the applicability of the classical
Maxwellian distribution to solar coronal plasma has accumulated. Perhaps, the most direct indication of the
Maxwellian distributions insufficiency is routine detection of the kappa-distributions in the solar wind
plasma even during the most quiet periods (i.e., in quasi-stationary conditions), which may imply the
presence of such kappa-distributions at the corona, where the solar wind is launched \citep{Maksimovic_etal_1997}. These kappa-distributions, characterized by a temperature $T$ and index $\kappa$, have different indices in the slow and fast solar wind flows, originating in the normal corona and coronal holes respectively, implying that the ``equilibrium'' distributions in the normal corona and coronal holes can have accordingly different indices. In addition, a number of coronal observations seem to require a kappa-like or $n$-distribution, e.g., some EUV line enhancements during solar flares \citep{dufton84, anderson96, pinfield99, dzifcakova11}. In fact, the temperature diagnostics based on the UV and X-ray lines depend on the temperature, the density (emission measure), and the distribution type. Specifically, for the kappa-distribution (compared to the Maxwellian one) the filter responses to emission are broader functions of $T$, and their maxima are flatter, which may result in a systematic error in coronal temperature diagnostics \citep{Dudic_etal_2009, dudik12}. Coronal hard X-ray (HXR) emission spectra are often well fit by a kappa-distribution of the flaring plasma \citep{Kasparova_Karlicky_2009, Oka_etal_2013}.

Finally, the radio data on the GR emission from various active regions \citep[e.g.,][]{Uralov_etal_2006,Uralov_etal_2008,Nita_etal_2011} seem to favor harmonic numbers larger than the expected value of three \citep{Lee_2007}, which could imply a deviation of the electron equilibrium distribution from a Maxwellian.
Although the Maxwellian distribution is often assumed to be a true equilibrium distribution of the plasma
particles, this is not necessarily the case. In terms of thermodynamics, the equilibrium distribution can easily be derived in the form of a Maxwellian if one adopts the system to be extensive (e.g., the entropy of the system is the sum of entropies of its macroscopic parts). Microscopically, at the kinetic level, this same
distribution is derived from the kinetic equation under the assumption that the equilibrium is achieved via
close binary (e.g., Coulomb) collisions. Stated another way, the Maxwellian distribution is a natural
equilibrium state of a closed collisional system.

In a non-extensive thermodynamical system \citep[e.g.,][]{Tsallis_1988, Leubner_2002}, however, the system
entropy is no longer an additive measure and so is not equal to the partial entropy sum. Accordingly, the
equilibrium distributions are not unique, and under certain assumptions the kappa-distribution can represent
the true equilibrium solution. Microscopically, this non-extensivity means that far interactions (rather than
close binary collisions) play a dominant role in reaching the equilibrium distribution; a sustained heat or
particle flux, if present, can further complicate the equilibrium established in such an open, non-extensive
system. In particular, a moving plasma (e.g., in the presence of a mean DC electric field) can have a
distribution similar to $n$-distribution \citep{Karlicky_etal_2012, Karlicky_2012}.

The coronal plasma may be a good example of such open, potentially non-extensive plasma. Indeed,
there is a sustained energy flux from lower layers of the solar atmosphere into the chromosphere and corona,
and the coronal plasma is indeed only weakly collisional, so that distant wave-particle interactions play a role
that is often much more important than the binary Coulomb collisions. The radio emission in general does
depend (in addition to the magnetic field) on the distant collisions and wave-particle interactions, and so
offers a sensitive probe of such interactions. Therefore, remote sensing of the solar corona may be explicitly
dependent on this new, fundamental physics of the collisionless plasma.

Radio measurements, with their significant optical depth, are potentially most sensitive to the distribution type, so it would be wise to use the radio measurements to address the fundamental question of the equilibrium/quasi-stationary distribution of the collisionless plasma of the solar corona. With the microwave imaging spectroscopy available from Jansky Very Large Array (JVLA) and Expanded Owens Valley Solar Array (EOVSA), spatially resolved radio spectra of the requisite quality will be available for the first time to probe this question.

In this paper we develop analytical theory of the GR and free-free emission from two, currently the most popular, non-Maxwellian distributions---namely, the kappa- and $n$-distributions. Currently, the only available element of this theory is the free-free emission from kappa-distributions with integer $\kappa$ \citep{Chiuderi_Drago_2004}, which we further develop and incorporate into the general framework.

In this study we do not take into account a possible moderate anisotropy of the distribution imposed by the external
magnetic field. Although in the presence of strong magnetic field the anisotropy of the plasma distribution
can be rather strong, it cannot be too strong in the quasi-stationary case discussed here; otherwise, a number
of instabilities will develop and give rise to coherent radio emission easily detectable and recognizable in
observations.

\begin{figure*}
\includegraphics{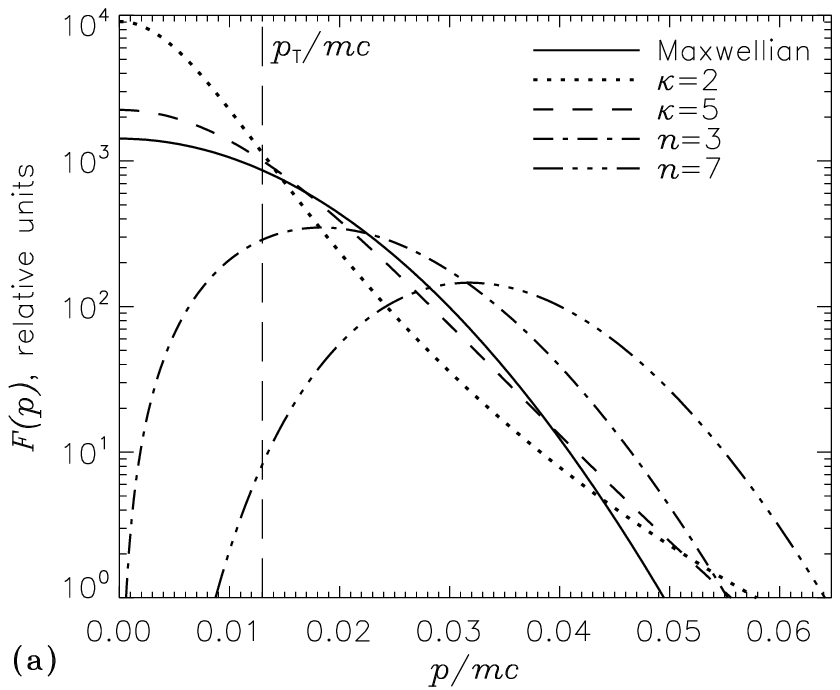}
\includegraphics{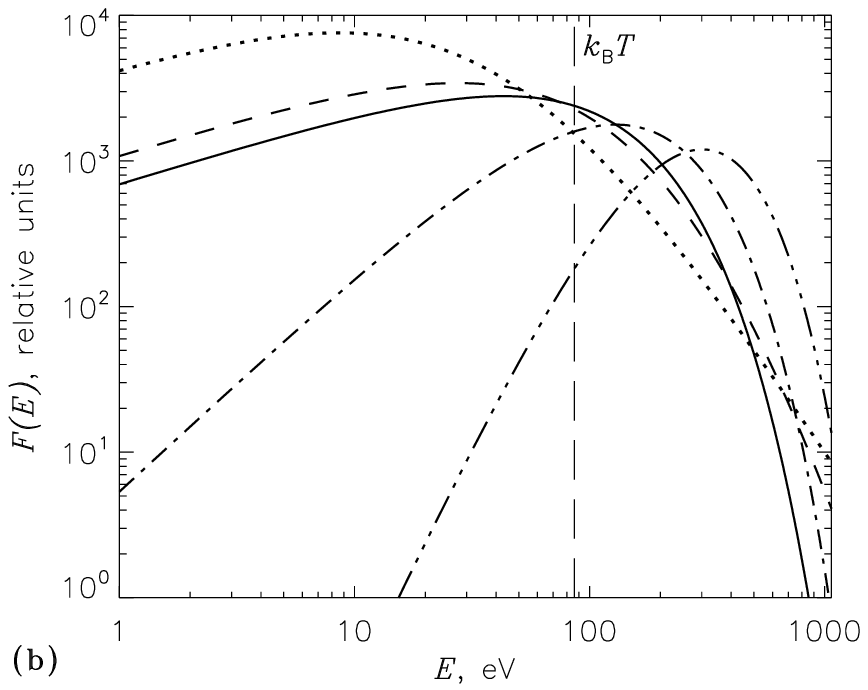}
\caption{Examples of the electron distribution functions (Maxwellian, kappa- and $n$-distribution) used in this paper. All plots were calculated for the plasma temperature of $T=10^6$ K. a) Distribution functions in the momentum space (see Section \protect\ref{S_Dis_Fun}); b) the same distribution functions in the energy space.}
\label{Fig_disFuns}
\end{figure*}

\section{Distribution functions}
\label{S_Dis_Fun}
In our study we will use the following distribution functions of the plasma particles, see Fig. \ref{Fig_disFuns}.

I. Maxwellian (thermal) distribution:
\begin{equation}\label{Eq_disfun_M_def}
      F_{\mathrm{M}}(\mathbf{p})=\frac{1}{(2\pi m k_{\mathrm{B}} T)^{3/2}}\exp\left\{a\frac{p^2}{2m k_{\mathrm{B}} T}\right\},
\end{equation}
where $a=-1$, $m$ is the electron mass,  $k_{\mathrm{B}}$ is the Boltzmann constant,  $T$ is the plasma temperature, and  $\mathbf{p}$ is the electron momentum vector.

II. $n-$distribution \citep{hares79, seely87, kulinova11, Karlicky_etal_2012, Karlicky_2012}:
\begin{equation}\label{Eq_disfun_n_def}
      F_{n}(\mathbf{p})=\frac{A_n}{(2\pi m k_{\mathrm{B}} T)^{3/2}}
      \left(\frac{p^2}{2 m k_{\mathrm{B}} T}\right)^{(n-1)/2}
      \exp\left\{-\frac{p^2}{2m k_{\mathrm{B}} T}\right\},
\end{equation}
where
\begin{equation}\label{Eq_disfun_n_An}
      A_n = \frac{\sqrt{\pi}}{2 \Gamma(n/2+1)}.
\end{equation}
Note that for all integer $l=(n-1)/2$ (i.e., for all odd $n=3, 5, ...$) the $n-$distribution can be obtained from the Maxwellian distribution by differentiating the latter by parameter $a$ and then accepting $a=-1$:

\begin{equation}\label{Eq_disfun_n_der_a}
      F_{n}(\mathbf{p})=A_n
      \frac{d^l}{da^l} F_{\mathrm{M}}(\mathbf{p})\Big|_{a=-1}.
\end{equation}
Apparently, the $n-$distribution with $n=1$ is equivalent to the Maxwellian one.

III. Kappa-distribution \citep{vasyliunas68, owocki83, Maksimovic_etal_1997, livadiotis09, pierrard10}:
\begin{equation}\label{Eq_disfun_kap_def}
      F_{\kappa}(\mathbf{p})=\frac{A_\kappa}{(2\pi m k_{\mathrm{B}} T)^{3/2}}
      \left(1+\frac{p^2}{2 (\kappa-3/2) m k_{\mathrm{B}} T}\right)^{-\kappa-1},
\end{equation}
where
\begin{equation}\label{Eq_disfun_k_Ak}
      A_\kappa = \frac{\Gamma(\kappa+1)}{\Gamma(\kappa-1/2)(\kappa-3/2)^{3/2}}.
\end{equation}

\section{General approach}
General definitions for arbitrary incoherent emission process are
\begin{equation}\label{Eq_emis_gen_def}
j_{f}^{\sigma}= \int I_{\mathbf{n},f}^{\sigma}F(\mathbf{p})\,\mathrm{d}^3\mathbf{p}= \int I_{f}^{\sigma}F(\mathbf{p}) p^2\,\mathrm{d}p,
\end{equation}
where $j_{f}^{\sigma}$ is the volume emissivity of the wave-mode $\sigma$ at the frequency $f$,  $I_{\mathbf{n},f}^{\sigma}$ is the radiation power emitted by a single particle with a given momentum $\mathbf{p}$ per unit time, frequency, and element of the solid angle, $I_{f}^{\sigma}$ is the same measure integrated over the full solid angle
\begin{equation}\label{Eq_emis_f_n_link}
I_{f}^{\sigma} = \int I_{\mathbf{n},f}^{\sigma}\,\mathrm{d}\Omega,
\end{equation}
and $F(\mathbf{p})$ is the particle distribution function normalized by $\mathrm{d}^3\mathbf{p}$, as those defined in \S~\ref{S_Dis_Fun}. Note that the second equality in Eq.~(\ref{Eq_emis_gen_def}) implies the distribution isotropy. Similarly, for the absorption coefficient we have
\begin{equation}\label{Eq_abso_gen_def}
\varkappa^{\sigma}= -\frac{c^2}{n_\sigma^2 f^2}\int I_{\mathbf{n},f}^{\sigma}\frac{1}{v}\left[\frac{\partial F(\mathbf{p})}{\partial p}+ \frac{1-\mu^2}{\mu p}\frac{\partial F(\mathbf{p})}{\partial\mu}\right]\mathrm{d}^3\mathbf{p}=
 -\frac{c^2}{n_\sigma^2 f^2} \int I_{f}^{\sigma}\frac{\partial F(\mathbf{p})}{v\partial p} p^2\,\mathrm{d}p,
\end{equation}
where $n_{\sigma}$ is the refraction index of the wave-mode $\sigma$, $\mathbf{v}$ is the electron velocity, $\beta=v/c$, and $\mu=\cos\alpha$ is the cosine of electron pitch-angle $\alpha$. Again, the second equality in Eq.~(\ref{Eq_abso_gen_def}) implies the distribution isotropy.

Gyrosynchrotron emission (magnetobremsstrahlung) from a single electron with arbitrary energy is described by the formula (e.g., FT13, Eq 9.148)
\begin{equation}
\label{Eq_GS_singl_part_Int}
I_{\mathbf{n},f}^\sigma=\frac{2\pi e^2}{c}\frac{n_\sigma f^2}{1+T_\sigma^2} \sum\limits_{s=-\infty}^{\infty}
 \left[\frac{T_\sigma(\cos\theta-n_\sigma\beta\mu)+L_\sigma\sin\theta}{n_\sigma\sin\theta}J_s(\lambda)+
J'_s(\lambda)\beta\sqrt{1-\mu^2}\right]^2\times$$$$
\times \delta\left[f(1-n_\sigma\beta\mu\cos\theta)-
\frac{sf_{\mathrm{Be}}}{\gamma}\right]=\sum\limits_{s=-\infty}^{\infty}I_{\mathbf{n},f,s}^\sigma,
\end{equation}
where $e$ is the electron charge, $f_{\mathrm{Be}}=eB/(2\pi mc)$ is the electron cyclotron frequency, $B$ is the magnetic field strength, $\theta$ is the viewing angle (the angle between the wave vector and the magnetic field vector), $\gamma$ is the Lorenz factor, $J_s$ is the Bessel function, and the parameters $T_{\sigma}$ and $L_{\sigma}$ are the components of the wave polarization vector. 
The argument of the Bessel functions is 
\begin{equation}
\lambda=\frac{f}{f_{\mathrm{Be}}}\gamma n_{\sigma}\beta\sin\theta\sqrt{1-\mu^2}.
\end{equation}
Classical theory of the GR radiation from a nonrelativistic  plasma (where $\beta\ll 1$ and hence $\lambda\ll 1$) employs small argument expansion of the Bessel functions and their derivatives and keeping the first non-vanishing terms of this expansion only, which yields for any term $s$ of the series
\begin{equation}
\label{Eq_GS_singl_part_cycl}
I_{\mathbf{n},f,s}^\sigma=\frac{2\pi e^2}{c}\frac{n_\sigma f}{1+T_\sigma^2}
\frac{s^{2s} n_\sigma^{2s-2}\beta^{2s}\sin^{2s-2}\theta (1-\mu^2)^s}{2^{2s} (s!)^2}
 \left[T_\sigma\cos\theta+L_\sigma\sin\theta+1\right]^2\times$$$$
\times \delta\left[n_\sigma\beta\mu\cos\theta -\left(1-\frac{sf_{\mathrm{Be}}}{f}\right)
 \right],
\end{equation}

Regarding the free-free emission, the power of bremsstrahlung emitted by a single electron (FT13, Eq 9.267) is
\begin{equation}
\label{Eq_I_f_brem_nonre_4}
 I_{f}^\sigma=n_\sigma\frac{16\pi e^6 Z^2 n_i}{3 v m^2 c^3} \ln\Lambda_{\mathrm{C}},
\end{equation}
where $n_i$ is the number density of target ions with charge number $Z$ and $\ln\Lambda_{\mathrm{C}}$ is the Coulomb logarithm.

\section{Emissions produced by $n$-distributions}
Given that the $n$-distribution can be obtained from the Maxwellian using Eq.~(\ref{Eq_disfun_n_der_a}) we can first calculate the emissivities and absorption coefficients from the Maxwellian distribution written in form (\ref{Eq_disfun_M_def}) and then determine the wanted emissivities and absorption coefficients from the $n$-distribution differentiating the corresponding Maxwellian expressions over the $a$ parameter $(n-1)/2$ times.

\subsection{Gyroresonance emission from Maxwellian distributions}
\label{s_GR_Max}
Classical theory of the GR radiation from a Maxwellian plasma requires finding the emissivity and absorption coefficient defined by Eqs.~(\ref{Eq_emis_gen_def}) and (\ref{Eq_abso_gen_def}) with the Maxwellian distribution, Eq.~(\ref{Eq_disfun_M_def}). We can easily perform this standard derivation  for arbitrary negative $a$ (the cylindrical coordinates $p_{\bot},~p_{\|},~\varphi$  are the most convenient to use here), which yields for the emissivity
\begin{equation}\label{Eq_emis_Max_a}
j_{f,s}^{\mathrm{M},\sigma}=\frac{\sqrt{2\pi}e^2n_e f}{c}
       \left(\frac{k_{\mathrm{B}} T}{mc^2}\right)^{s-1/2}
       \frac{s^{2s}n_\sigma^{2s-2}\sin^{2s-2}\theta}{2^s s!(1+T_\sigma^2) |\cos\theta|}
       [T_\sigma\cos\theta + L_\sigma\sin\theta+1]^2$$$$
       \left(-\frac{1}{a}\right)^{s+1}
       \exp\left\{a \frac{mc^2}{2k_{\mathrm{B}} T}\frac{(f-sf_{\mathrm{Be}})^2}{f^2 n_\sigma^2\cos^2\theta} \right\},
\end{equation}
and absorption coefficient
\begin{equation}\label{Eq_abso_Max_a}
\varkappa_{s}^{\mathrm{M},\sigma}=\frac{\sqrt{2\pi}e^2n_e c}{f k_{\mathrm{B}} T}
       \left(\frac{k_{\mathrm{B}} T}{mc^2}\right)^{s-1/2}
       \frac{s^{2s}n_\sigma^{2s-4}\sin^{2s-2}\theta}{2^s s!(1+T_\sigma^2) |\cos\theta|}
       [T_\sigma\cos\theta + L_\sigma\sin\theta+1]^2$$$$
       \left(-\frac{1}{a}\right)^{s}
       \exp\left\{a \frac{mc^2}{2k_{\mathrm{B}} T}\frac{(f-sf_{\mathrm{Be}})^2}{f^2 n_\sigma^2\cos^2\theta} \right\}.
\end{equation}

It is straightforward to check that for $a=-1$ the source function $S_{f,s}^\sigma=j_{f,s}^\sigma/\varkappa_{s}^\sigma$ obeys Kirchhoff's law as required:
\begin{equation}\label{Eq_Kirchh_Max}
S_{f,s}^\sigma=\frac{j_{f,s}^\sigma}{\varkappa_{s}^\sigma}=\frac{n_\sigma^{2}f^2}{c^2} k_{\mathrm{B}} T.
\end{equation}

To obtain final formulae of the GR emission from a nonuniform source (nonuniform magnetic field at first place) we have yet to integrate the equations obtained over the resonance layer. To do so we expand the spatial dependence of the magnetic field around the resonance value:
\begin{equation}\label{Eq_B_z_expan}
 B(z)\approx B_0 \left(1+\frac{z}{L_{\mathrm{B}}}\right),
\end{equation}
where $B_0=2\pi f mc/(s e)$ is the resonant value of the magnetic field for the frequency $f$ at the harmonic $s$, $z$ is the spatial coordinate along the line of sight with $z=0$ at $B=B_0$, and
\begin{equation}\label{Eq_L_B}
L_{\mathrm{B}}=\left(\frac{1}{B}\frac{\partial B}{\partial z}\right)^{-1}.
\end{equation}

With these definitions we get $sf_{\mathrm{Be}}=f(1+z/L_{\mathrm{B}})$, so the exponent reads
\begin{equation}\label{Eq_expo_term}
       \exp\left\{a \frac{mc^2}{2k_{\mathrm{B}} T}\frac{(f-sf_{\mathrm{Be}})^2}{f^2 n_\sigma^2\cos^2\theta} \right\}\approx
       \exp\left\{a \frac{mc^2}{2k_{\mathrm{B}} T}\frac{z^2}{L_{\mathrm{B}}^2 n_\sigma^2\cos^2\theta} \right\}.
\end{equation}
Now we can find the optical depth of the $s$-th gyrolayer by integrating the absorption coefficient along the line of sight:
\begin{equation}\label{Eq_tau_Max_a}
 \tau_{s}^{\mathrm{M},\sigma}=\int\limits_{-\infty}^{\infty}\varkappa_{s}^{\mathrm{M},\sigma}(z) dz=
        \frac{\pi e^2n_e }{f mc}
       \left(\frac{k_{\mathrm{B}} T}{mc^2}\right)^{s-1}
       \frac{s^{2s}n_\sigma^{2s-3}\sin^{2s-2}\theta}{2^{s-1} s!(1+T_\sigma^2) } L_{\mathrm{B}}[T_\sigma\cos\theta + L_\sigma\sin\theta+1]^2
       \left(-\frac{1}{a}\right)^{s+1/2}
\end{equation}
and, accordingly, the emissivity along the line of sight:
\begin{equation}\label{Eq_emisZ_Max_a}
 J_{f,s}^{\mathrm{M},\sigma}=\int\limits_{-\infty}^{\infty}j_{f,s}^{\mathrm{M},\sigma} (z) dz=
        \frac{\pi e^2n_e f}{c}
       \left(\frac{k_{\mathrm{B}} T}{mc^2}\right)^{s}
       \frac{s^{2s}n_\sigma^{2s-1}\sin^{2s-2}\theta}{2^{s-1} s!(1+T_\sigma^2) } L_{\mathrm{B}}       [T_\sigma\cos\theta + L_\sigma\sin\theta+1]^2
       \left(-\frac{1}{a}\right)^{s+3/2}.
\end{equation}
Here we assume that the dependence of the emissivity and absorption coefficient on the coordinate $z$ is only caused by  exponent (\ref{Eq_expo_term}), while all other factors in  expressions (\ref{Eq_emis_Max_a}) and (\ref{Eq_abso_Max_a}) are approximately constant within the gyrolayer. Apparently,  the obtained expressions coincide with classical GR formulae \citep[e.g.,][]{Zheleznyakov_1970} for $a=-1$ and obey Kirchhoff's law (\ref{Eq_Kirchh_Max}); in particular
\begin{equation}\label{Eq_Kirchh_Max_Jtau}
\frac{J_{f,s}^\sigma}{\tau_{s}^\sigma}=\frac{j_{f,s}^\sigma}{\varkappa_{s}^\sigma}=S_{f,s}^\sigma=\frac{n_\sigma^{2}f^2}{c^2} k_{\mathrm{B}} T.
\end{equation}
Thus, the GR emission intensity from a given gyrolayer can be written down simply as
\begin{equation}\label{Eq_Inten_GR_layer}
{\cal J}_{f,s}^\sigma=S_{f,s}^\sigma\left[1-\exp(-\tau_{s}^\sigma)\right]= \frac{J_{f,s}^\sigma}{\tau_{s}^\sigma}\left[1-\exp(-\tau_{s}^\sigma)\right],
\end{equation}
using the measures integrated over the gyrolayer, which simplifies the theory greatly.

\subsection{Gyroresonance emission from $n$-distributions}
Using Eq.~(\ref{Eq_disfun_n_der_a}) we can immediately write down the GR formulae for the $n$-distribution (for odd values of $n$):
\begin{equation}\label{Eq_GR_n_gen}
      \Phi^{(n)}=\left.A_n
      \frac{d^l}{da^l} \Phi^{(\mathrm{M})}\right|_{a=-1},
\end{equation}
where $\Phi^{(n)}$ is any of $j_{f,s}^{\sigma}$, $\varkappa_{s}^{\sigma}$, $J_{f,s}^{\sigma}$, and $\tau_{s}^{\sigma}$ for $n$-distribution and $\Phi^{(\mathrm{M})}$ is the corresponding measure for the Maxwellian plasma. As a result, general expressions for the emissivity and absorption coefficient take the form
\begin{equation}\label{j_n}
j^{n, \sigma}_{f, s}=A_nj^{\mathrm{M}, \sigma}_{f, s}\sum\limits_{q=0}^l\frac{l!}{s!}\frac{(s+l-q)!}{(l-q)!q!}\left(\frac{\zeta_s^2}{2}\right)^q,
\end{equation}
\begin{equation}\label{k_n}
\varkappa^{n, \sigma}_s=A_n\varkappa^{\mathrm{M}, \sigma}_s\sum\limits_{q=0}^l\frac{l!}{(s-1)!}\frac{(s-1+l-q)!}{(l-q)!q!}\left(\frac{\zeta_s^2}{2}\right)^q,
\end{equation}
where $l=(n-1)/2$ and
\begin{equation}\label{Eq_zeta}
\zeta_{s}^2=\frac{mc^2}{k_{\mathrm{B}} T}\frac{(f-sf_{\mathrm{Be}})^2}{f^2 n_\sigma^2\cos^2\theta}=
\frac{\beta_z^2}{\beta_{\mathrm{T}}^2},
\end{equation}
\begin{equation}
\beta_z=\frac{f-sf_{\mathrm{Be}}}{f n_\sigma |\cos\theta|},\qquad
\beta_{\mathrm{T}}^2=\frac{k_{\mathrm{B}} T}{mc^2}.
\end{equation}

For example for $n=3$ we obtain
\begin{equation}\label{Eq_emis_n_3}
j_{f,s}^{(3),\sigma}=A_3 j_{f,s}^{\mathrm{M},\sigma}\left(s+1+\frac{\zeta_s^2}{2}\right),\quad
\varkappa_{s}^{(3),\sigma}=A_3 \varkappa_{s}^{\mathrm{M},\sigma}\left(s+\frac{\zeta_s^2}{2}\right).
\end{equation}
Similarly, for $n=5$ we obtain:
\begin{equation}\label{Eq_emis_n_5}
j_{f,s}^{(5),\sigma}=A_5 j_{f,s}^{\mathrm{M},\sigma}\left[(s+1)(s+2)+(s+1)\zeta_s^2+\frac{\zeta_s^4}{4}\right],\quad
\varkappa_{s}^{(5),\sigma}=A_5\varkappa_{s}^{\mathrm{M},\sigma}\left[s(s+1)+s\zeta_s^2+\frac{\zeta_s^4}{4}\right].
\end{equation}

It is easy to see that, unlike  the Maxwellian case, the source function, Eq.~(\ref{Eq_Kirchh_Max}), does not take place any longer for the $n$-distributions; moreover, in addition to the standard frequency dependence $\propto f^2$, the source function also depends on $\zeta_s$ and on the harmonic number $s$.  Kirchhoff's law recovers only for $\zeta_s\gg 1$, i.e., outside the GR layer, where the GR emissivity and opacity are both exponentially small. Note that in spite of the positive derivative of the $n$-distribution over energy (or momentum modulus) the GR absorption coefficient  (in the nonrelativistic approximation) is always positive, so no electron-cyclotron maser instability takes place for the isotropic $n$-distributions.

To obtain the optical depth and emissivity integrated along the line of sight, one can integrate expressions (\ref{j_n}--\ref{k_n}) 
in the way suggested by Eqs.~(\ref{Eq_tau_Max_a}) and (\ref{Eq_emisZ_Max_a}). A more practical way, however, is to apply Eq.~(\ref{Eq_GR_n_gen}) to $J_{f,s}^{\sigma}$, and $\tau_{s}^{\sigma}$ directly, which yields
\begin{equation}\label{Eq_emisZ_n}
J_{f,s}^{(n),\sigma}=A_n  \frac{(s+n/2)\Gamma(s+n/2)}{(s+1/2)\Gamma(s+1/2)} J_{f,s}^{\mathrm{M},\sigma},
\end{equation}
\begin{equation}\label{Eq_tau_n}
\tau_{s}^{(n),\sigma}=A_n  \frac{\Gamma(s+n/2)}{\Gamma(s+1/2)} \tau_{s}^{\mathrm{M},\sigma},
\end{equation}
which is here written in the form applicable to arbitrary $n$---not necessarily the odd integer numbers.
The Kirchhoff's law ``generalization'' to the GR emission from $n$-distributions reads:
\begin{equation}\label{Eq_Kirchh_n_GR}
\overline{S_{f,s}^{(n),\sigma}}=\frac{J_{f,s}^{(n),\sigma}}{\tau_{s}^{(n),\sigma}}=\frac{(s+n/2)}{(s+1/2)}\frac{n_\sigma^{2}f^2}{c^2} k_{\mathrm{B}} T.
\end{equation}
This equation converges to usual Kirchhoff's law for $n=1$ as required since $n=1$ means the Maxwellian distribution. Then, it also approaches the usual Kirchhoff's law for large $s\gg n/2$. The reason is that the higher gyroharmonics are produced by more energetic electrons from the distribution tails, which are similar for both Maxwellian and $n$-distributions.

We emphasize that the local source function, $S_{f,s}^\sigma=j_{f,s}^{\sigma}/\varkappa_{s}^{\sigma}$, which is now a function of the coordinate $z$,  is no longer equal to the averaged one $\overline{S_{f,s}^{(n),\sigma}}$ unlike in the Maxwellian case. This further implies that the GR intensity from a given gyrolayer cannot be written in simple form (\ref{Eq_Inten_GR_layer}), but requires more exact knowledge of the source function value at  the level (inside the gyrolayer) making the dominant contribution to the intensity. Inspection of expressions (\ref{Eq_emis_n_3}) or (\ref{Eq_emis_n_5}) suggests that in the optically thick gyrolayer the radiation intensity will \textit{decrease} with the optical depth \textit{increase}. We return to this point later, in \S~\ref{s_transfer}.

\begin{figure*}
\includegraphics{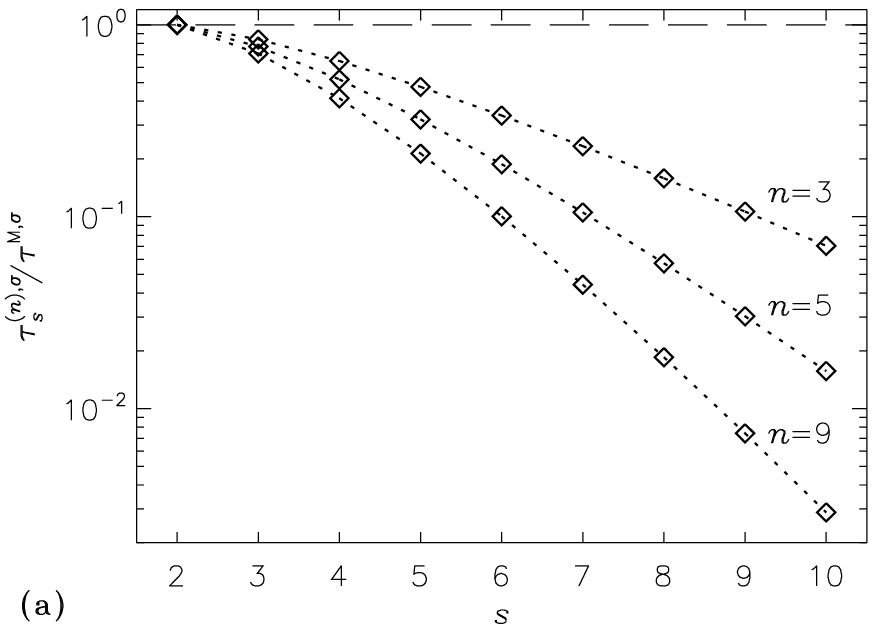}
\includegraphics{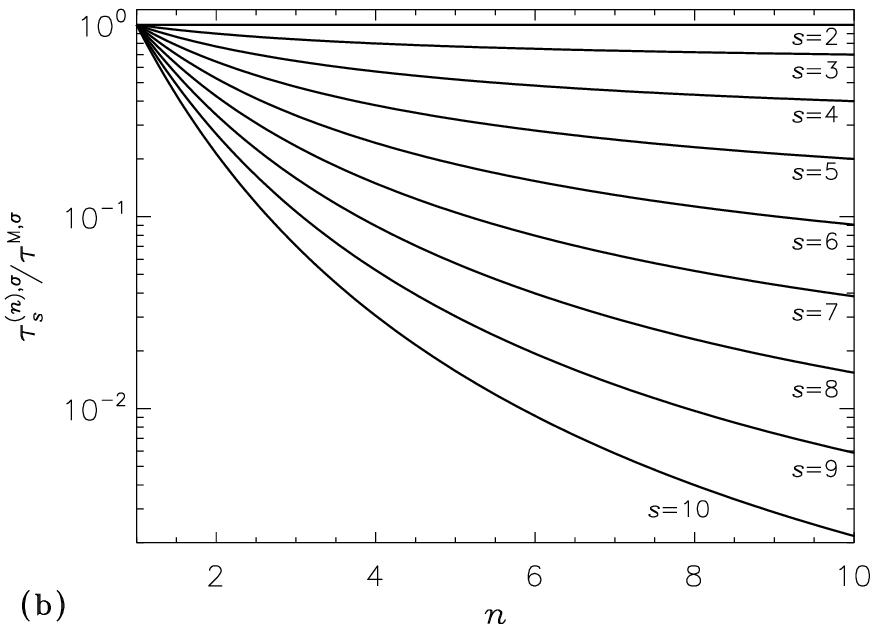}
\caption{Ratio of the optical depths of gyroresonance layers for the $n$-distribution to the Maxwellian ones. a) $\tau^{(n), \sigma}_s/\tau^{\mathrm{M}, \sigma}_s$ vs. $s$ for different $n$-indices. b) $\tau^{(n), \sigma}_s/\tau^{\mathrm{M}, \sigma}_s$ vs. $n$ for different harmonic numbers.}
\label{FigTauN}
\end{figure*}

Figure \ref{FigTauN} demonstrates the ratio of optical depths of gyroresonance layers for the $n$- and Maxwellian distributions, according to Eq. (\ref{Eq_tau_n}); this ratio depends only on $s$ and $n$. We should note that the parameter $T$ does not play a role of the effective energy for the $n$-distribution. The ``pseudo-temperature'' $T_*=T(n+2)/3$ computed as the second moment of the distribution is a true measure of the average electron energy \citep{dzifcakova98, dzifcakova01, dzifcakova08}; in Fig. \ref{FigTauN}, the ``pseudo-temperature'' $T_*$ (instead of $T$) is assumed to be the same for all distributions, which results in an additional correction factor of $[3/(n+2)]^{s-1}$ in the right side of Eq. (\ref{Eq_tau_n}). We can see that the optical depth of a gyrolayer for the $n$-distributions, in general, is \textit{smaller} than that for the Maxwellian distribution; the ratio of optical depths decreases with the increase of the harmonic number and/or the $n$-index. It is interesting to note that for the second gyrolayer, the optical depth is exactly the same for the Maxwellian and $n$-distributions with arbitrary $n$-index.

\begin{figure*}
\includegraphics{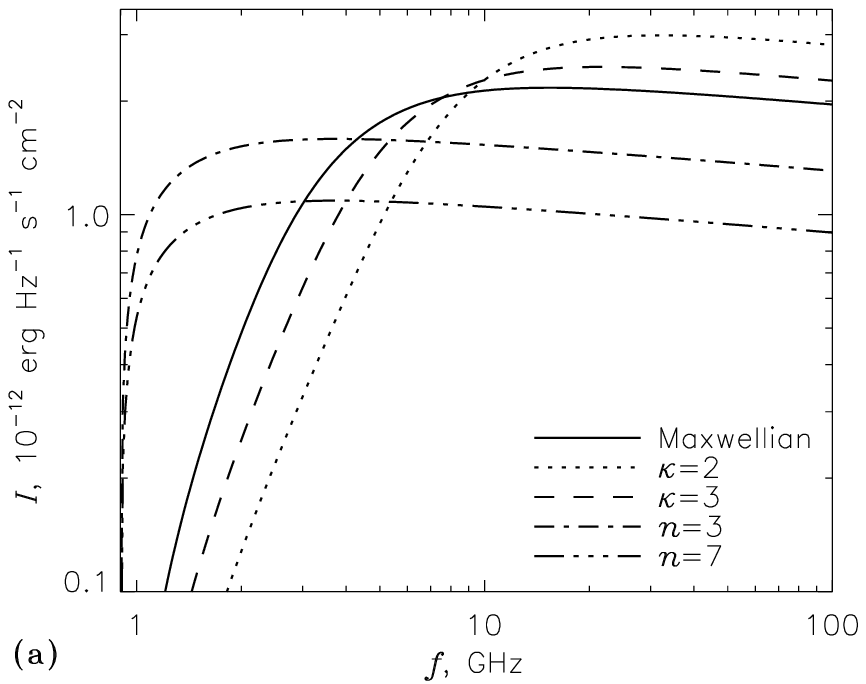}
\includegraphics{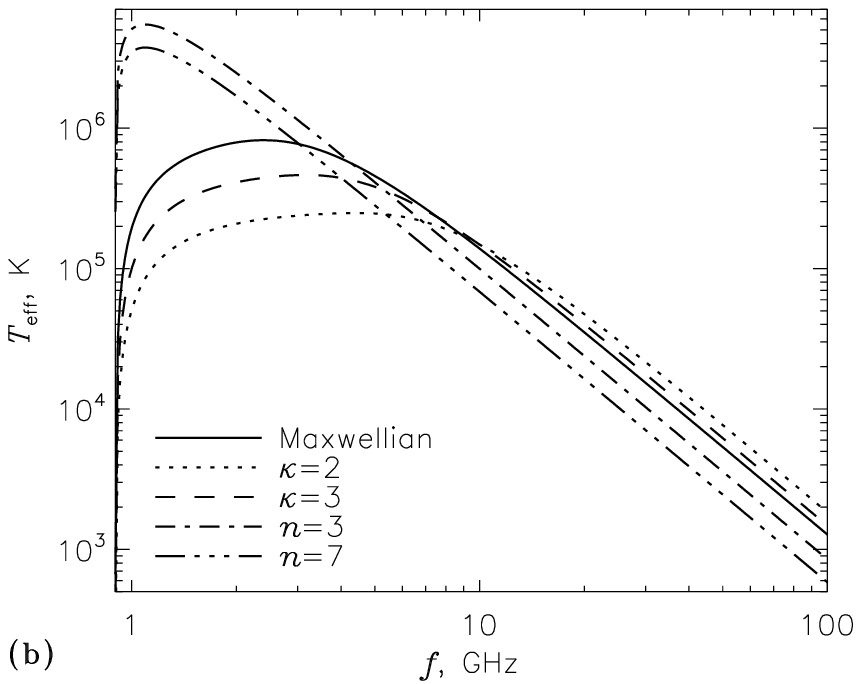}
\caption{Intensity spectra (a) and brightness temperatures (b) of the free-free emission from the Maxwellian, kappa- and $n$-distributions. The emission parameters were computed for a fully ionized unmagnetized hydrogen plasma with the density of $n_0=10^{10}$ $\textrm{cm}^{-3}$  (that corresponds to the plasma frequency of about 0.9 GHz) and temperature of $T=10^6$ K; the source size along the line-of-sight is $L=10^9$ cm.}
\label{FigFF}
\end{figure*}

\subsection{Free-free emission from Maxwellian distributions}
To compute the free-free emission from the Maxwellian distribution is even easier than the GR emission considered above.
We can discard the weak dependence of the Coulomb logarithm on the particle energy  to the first approximation while taking integrals in Eqs.~(\ref{Eq_emis_gen_def}) and (\ref{Eq_abso_gen_def}). These integrations are straightforward; they yield well-known results for the emissivity
\begin{equation}\label{Eq_emis_ff_Max_a}
j_{f,\mathrm{ff}}^{\mathrm{M},\sigma}=-\frac{8e^6  n_\sigma n_e n_i \ln\Lambda_{\mathrm{C}}}{3 \sqrt{2\pi} (m c^2)^{3/2}(k_{\mathrm{B}} T)^{1/2}}
       \frac{1}{a}; \qquad a=-1,
\end{equation}
and absorption coefficient
\begin{equation}\label{Eq_abso_ff_Max_a}
\varkappa_{\mathrm{ff}}^{\mathrm{M},\sigma}= \frac{8e^6   n_e n_i \ln\Lambda_{\mathrm{C}}}{3 \sqrt{2\pi}n_\sigma  c f^2 (m k_{\mathrm{B}} T)^{3/2}}.
\end{equation}
Evidently, these expressions obey Kirchhoff's law as needed for the thermal emissions.

\subsection{Free-free emission from $n$-distributions}
Now the theory of the free-free emission from the $n$-distributions is derived from that for the Maxwellian distribution by consecutive  differentiating the obtained emission and absorbtion coefficients over $a$ parameter. For the emissivity we obtain
\begin{equation}\label{Eq_emis_ff_Max_l}
j_{f,\mathrm{ff}}^{(n),\sigma}=A_n l! \frac{8e^6  n_\sigma n_e n_i \ln\Lambda_{\mathrm{C}}}{3 \sqrt{2\pi} (m c^2)^{3/2}(k_{\mathrm{B}} T)^{1/2}};
\qquad l=(n-1)/2,
\end{equation}
which allows a straightforward analytical continuation to a non-integer $l$:
\begin{equation}\label{Eq_emis_ff_Max_any}
j_{f,\mathrm{ff}}^{(n),\sigma}= \frac{\sqrt{\pi} \Gamma(n/2+1/2)}{2\Gamma(n/2+1)}\frac{8e^6  n_\sigma n_e n_i \ln\Lambda_{\mathrm{C}}}{3 \sqrt{2\pi} (m c^2)^{3/2}(k_{\mathrm{B}} T)^{1/2}},
\end{equation}
where expression (\ref{Eq_disfun_n_An}) for the normalization constant $A_n$ has been taken into account. It is easy to estimate that the free-free emissivity slightly decreases compared with the Maxwellian distribution with the same $T$ parameter as $n$ increases.

Unlike the emissivity, the absorption coefficient described by Eq.~(\ref{Eq_abso_ff_Max_a}) does not depend on $a$; thus, all derivatives of the absorption coefficients over  this parameter are zeros, which means no free-free absorption by electrons with the $n$-distribution. This happens because the positive contribution to the absorption coefficient from the negative slope of this distribution at high velocities is fully compensated by the negative contribution (amplification) from the positive slope at low velocities. This corresponds to a marginal stability state when a non-zero emissivity is accompanied by zero absorption coefficient. No analogy to Kirchhoff's law can be formulated in this case; arbitrarily deep plasma with such a distribution remains optically thin as is clearly seen from Fig.~\ref{FigFF}.

\section{Emissions produced by kappa-distribution}
\subsection{Gyroresonance emission from kappa-distribution}
Integrals~(\ref{Eq_emis_gen_def}) and (\ref{Eq_abso_gen_def}) with the cyclotron radiation power, Eq.~(\ref{Eq_GS_singl_part_cycl}), are convenient to take in the cylindrical coordinates: integration over the azimuth angle results in $2\pi$ factor, while the integral over $dp_\|$ is taken with the $\delta$-function. The remaining single integration over $dp_{\bot}^2$ is a tabular integral of a rational fraction, which yields the emissivity
\begin{equation}\label{Eq_emis_kappa}
j_{f,s}^{\kappa,\sigma}=\frac{\sqrt{2\pi}e^2n_e f}{c}\frac{(\kappa-3/2)^{s-1/2}\Gamma(\kappa-s)}{\Gamma(\kappa-1/2)}
       \left(\frac{k_{\mathrm{B}} T}{mc^2}\right)^{s-1/2}
       \frac{s^{2s}n_\sigma^{2s-2}\sin^{2s-2}\theta}{2^s s!(1+T_\sigma^2) |\cos\theta|}\times$$$$
       \frac{[T_\sigma\cos\theta + L_\sigma\sin\theta+1]^2}
       {\left[1+ \frac{\zeta_s^2}{2(\kappa-3/2)} \right]^{\kappa-s}},\qquad
\kappa>s,			
\end{equation}
and the absorption coefficient
\begin{equation}\label{Eq_abso_kappa}
\varkappa_{s}^{\kappa,\sigma}=\frac{\sqrt{2\pi}e^2n_e c}{f k_{\mathrm{B}} T}
        \frac{(\kappa-3/2)^{s-3/2}(\kappa-s)\Gamma(\kappa-s)}{\Gamma(\kappa-1/2)}
       \left(\frac{k_{\mathrm{B}} T}{mc^2}\right)^{s-1/2}
       \frac{s^{2s}n_\sigma^{2s-4}\sin^{2s-2}\theta}{2^s s!(1+T_\sigma^2) |\cos\theta|}\times$$$$
       \frac{[T_\sigma\cos\theta + L_\sigma\sin\theta+1]^2} 
       {\left[1+ \frac{\zeta_s^2}{2(\kappa-3/2)} \right]^{\kappa-s+1}},\qquad
\kappa>s-1,			
\end{equation}
where the parameter $\zeta_s$ is defined by Eq.~(\ref{Eq_zeta}). The source function
\begin{equation}\label{Eq_Kirchh_kappa_GR}
S_{f,s}^{\kappa,\sigma}=\frac{j_{f,s}^{\kappa,\sigma}}{\varkappa_{s}^{\kappa,\sigma}}=\frac{(\kappa-3/2)}{(\kappa-s)}\frac{n_\sigma^{2}f^2}{c^2} k_{\mathrm{B}} T
\left[1+ \frac{\zeta_s^2}{2(\kappa-3/2)} \right]
\end{equation}
depends on the $\zeta_s$ parameter that complicates the GR theory significantly for the same reason that has been explained for the $n$-distribution.
However, unlike $n$-distribution, here the GR intensity from an optically thick gyrolayer increases as the optical depth increases.

Now we can find the optical depth of the $s$-th gyrolayer by integrating the absorption coefficient in the linearly changing magnetic field, Eq.~(\ref{Eq_B_z_expan}), along the line of sight:
\begin{equation}\label{Eq_tau_kappa}
 \tau_{s}^{\kappa,\sigma}=\int\limits_{-\infty}^{\infty}\varkappa_{s}^{\kappa,\sigma}(z)\,\mathrm{d}z=
        \frac{\pi e^2n_e }{f mc}
        \frac{(\kappa-3/2)^{s-1}\Gamma(\kappa-s+1/2)}{\Gamma(\kappa-1/2)}
       \left(\frac{k_{\mathrm{B}} T}{mc^2}\right)^{s-1}
       \frac{s^{2s}n_\sigma^{2s-3}\sin^{2s-2}\theta}{2^{s-1} s!(1+T_\sigma^2) } L_{\mathrm{B}}\times$$$$
       [T_\sigma\cos\theta + L_\sigma\sin\theta+1]^2,\qquad
\kappa>s-1/2,			
\end{equation}
and, accordingly, the emissivity along the line of sight:
\begin{equation}\label{Eq_emisZ_kappa}
 J_{f,s}^{\kappa,\sigma}=\int\limits_{-\infty}^{\infty}j_{f,s}^{\kappa,\sigma} (z)\,\mathrm{d}z=
        \frac{\pi e^2n_e f}{c}
        \frac{(\kappa-3/2)^{s}\Gamma(\kappa-s-1/2)}{\Gamma(\kappa-1/2)}
       \left(\frac{k_{\mathrm{B}} T}{mc^2}\right)^{s}
       \frac{s^{2s}n_\sigma^{2s-1}\sin^{2s-2}\theta}{2^{s-1} s!(1+T_\sigma^2) } L_{\mathrm{B}}\times$$$$
       [T_\sigma\cos\theta + L_\sigma\sin\theta+1]^2,\qquad
\kappa>s+1/2.
\end{equation}
The ratio of these two expressions yields the effective source function (again, different from the Maxwellian's one):
\begin{equation}\label{Eq_Kirchh_kappaZ_GR}
\overline{S_{f,s}^{\kappa,\sigma}}=\frac{J_{f,s}^{\kappa,\sigma}}{\tau_{s}^{\kappa,\sigma}}= \frac{(\kappa-3/2)}{(\kappa-s-1/2)}\frac{n_\sigma^{2}f^2}{c^2} k_{\mathrm{B}} T.
\end{equation}
Since  kappa-distribution (\ref{Eq_disfun_kap_def}--\ref{Eq_disfun_k_Ak}) converges to the Maxwellian one when $\kappa\to\infty$, the GR emission parameters for large $\kappa$-indices ($\kappa\gg s$) approach those for the Maxwellian distribution; in particular,  relation (\ref{Eq_Kirchh_kappaZ_GR}) approaches the usual Kirchhoff's law.

Note that the above equations are only valid for relatively small gyroharmonics (otherwise, the corresponding integrals diverge), so that the derived here GR theory for the kappa-distribution may only be applicable  at $s<\kappa-1/2$. For higher harmonics, $s>\kappa-1/2$, the quasi-continuum gyrosynchrotron contribution from the power-law tail of the kappa-distribution, where the non-relativistic expansions used above are invalid,  dominates over the contribution from the nonrelativistic core of the distribution. If needed, this contribution can be computed in a usual way \citep{Fl_Kuzn_2010}.

\begin{figure}
\includegraphics{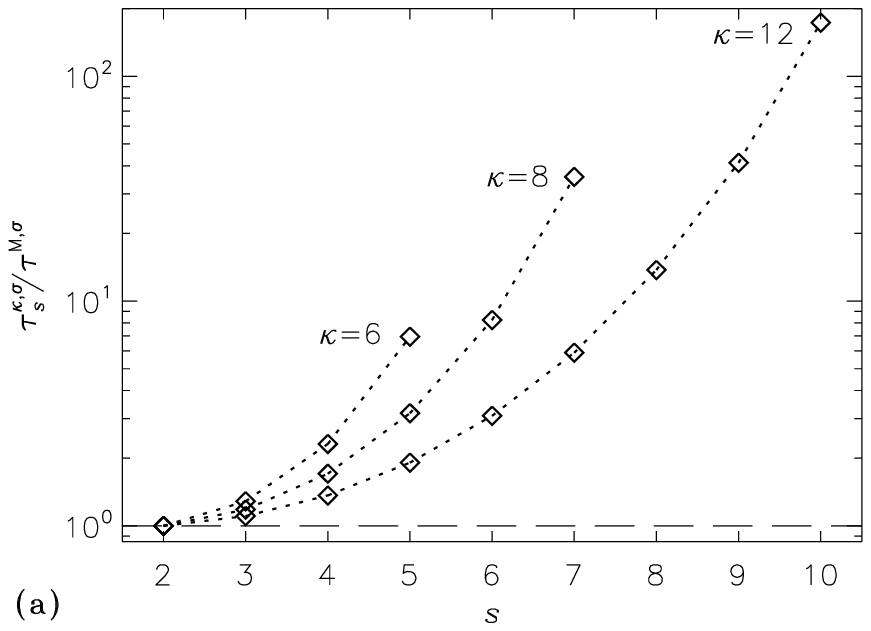}
\includegraphics{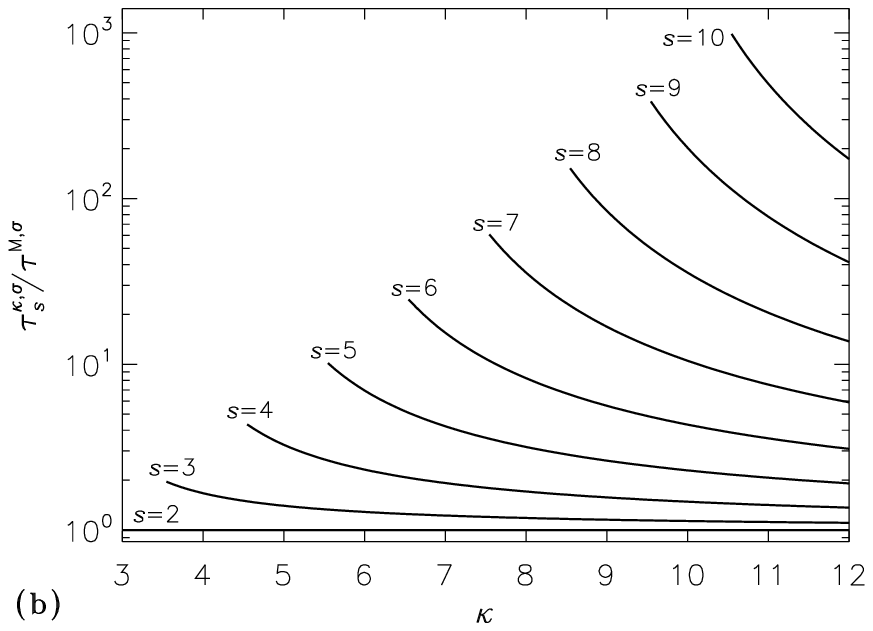}
\caption{Ratio of the optical depths of gyroresonance layers for the kappa-distribution to the Maxwellian ones.
Only finite-range data is presented because for each finite $\kappa$ there is a highest harmonic number, $s<\kappa-1/2$, up to which the developed theory is valid.
a) $\tau^{\kappa, \sigma}_s/\tau^{\mathrm{M}, \sigma}_s$ vs. $s$ for different $\kappa$-indices. b) $\tau^{\kappa, \sigma}_s/\tau^{\mathrm{M}, \sigma}_s$ vs. $\kappa$ for different harmonic numbers.}
\label{FigTauKappa}
\end{figure}

Figure \ref{FigTauKappa} demonstrates the ratio of optical depths of gyroresonance layers for the kappa- and Maxwellian distributions, according to Eqs. (\ref{Eq_tau_kappa}) and (\ref{Eq_tau_Max_a}); this ratio depends only on $s$ and $\kappa$. We can see that the optical depth of a gyrolayer for the kappa-distributions, in general, is \textit{larger} than that for the Maxwellian distribution; the ratio of optical depths increases with the increase of the harmonic number and/or  decrease of the kappa-index. The optical depth of the second gyrolayer is the same for all considered distributions---the Maxwellian, $n$- and kappa-distributions. The reason for this equivalence is that all these optical depths are proportional to $T^{s-1}$; thus, linearly proportional to $T$ for $s=2$. This means that the optical depths of the second gyrolayer are defined by the second moment of the given distribution only, that is the mean electron energy, which is adopted the same for all these distributions.

\subsection{Free-free emission from kappa-distribution}
\citet{Chiuderi_Drago_2004} developed analytical theory of free-free emission from kappa-distributions with integer indices $\kappa$. It is straightforward, however, to extend this theory to arbitrary real index $\kappa$. To do so we consider again Eqs.~(\ref{Eq_emis_gen_def}) and (\ref{Eq_abso_gen_def}) with the free-free radiation power, Eq.~(\ref{Eq_I_f_brem_nonre_4}),  but with kappa-distribution (\ref{Eq_disfun_kap_def}) instead of the Maxwellian one. Neglecting the (weak) energy dependence of the Coulomb logarithm as before, we can easily take the remaining integrals which yields for the emissivity
\begin{equation}\label{Eq_emis_ff_kappa}
j_{f,\mathrm{ff}}^{\kappa,\sigma}=A_\kappa \frac{\kappa-3/2}{\kappa}\frac{8e^6  n_\sigma n_e n_i \ln\Lambda_{\mathrm{C}}}{3 \sqrt{2\pi} (m c^2)^{3/2}(k_{\mathrm{B}} T)^{1/2}},
\end{equation}
and absorption coefficient
\begin{equation}\label{Eq_abso_ff_kappa}
\varkappa_{\mathrm{ff}}^{\kappa,\sigma}=A_\kappa \frac{8e^6 n_e n_i \ln\Lambda_{\mathrm{C}}}{3 \sqrt{2\pi}n_\sigma  c f^2 (m k_{\mathrm{B}} T)^{3/2}}.
\end{equation}

\citet{Chiuderi_Drago_2004} took into account the energy dependence of the Coulomb logarithm, which allowed them to obtain the results in the closed form for integer $\kappa$ only. This results in small corrections to the Coulomb logarithm, slightly different for the emissivity and absorption coefficient. With these corrections, which we interpolated with the parenthetical expressions below,  we can write
\begin{equation}\label{Eq_emis_ff_kappa_corr}
j_{f,\mathrm{ff}}^{\kappa,\sigma}=A_\kappa \frac{\kappa-3/2}{\kappa}\frac{8e^6  n_\sigma n_e n_i \ln\Lambda_{\mathrm{C}}}{3 \sqrt{2\pi} (m c^2)^{3/2}(k_{\mathrm{B}} T)^{1/2}}\left[1 - \frac{0.525(4/\kappa)^{1.25}}{\ln\Lambda_{\mathrm{C}}} \right],
\end{equation}
and
\begin{equation}\label{Eq_abso_ff_kappa_corr}
\varkappa_{\mathrm{ff}}^{\kappa,\sigma}=A_\kappa \frac{8e^6   n_e n_i \ln\Lambda_{\mathrm{C}}}{3 \sqrt{2\pi}n_\sigma  c f^2 (m k_{\mathrm{B}} T)^{3/2}}
    \left[1 - \frac{0.575(6/\kappa)^{1.1}}{\ln\Lambda_{\mathrm{C}}} \right].
\end{equation}

Therefore, Kirchhoff's law extension to the free-free emission from the kappa-distribution reads
\begin{equation}\label{Eq_Kirchh_kappa_ff}
S_{f,\mathrm{ff}}^{\kappa,\sigma}=\frac{j_{f,\mathrm{ff}}^{\kappa,\sigma}}{\varkappa_{\mathrm{ff}}^{\kappa,\sigma}}\approx \frac{\kappa-3/2}{\kappa}\frac{n_\sigma^{2}f^2}{c^2} k_{\mathrm{B}} T,
\end{equation}
where we discarded the ratio of two parenthetical expressions entering Eqs.~(\ref{Eq_emis_ff_kappa_corr}) and (\ref{Eq_abso_ff_kappa_corr}), which are both close to one, for brevity. Eq.~(\ref{Eq_Kirchh_kappa_ff}) implies that the effective  temperature from a plasma volume with kappa-distribution is lower than that for the Maxwellian plasma with the same temperature $T$; \citep[see][for greater detail]{Chiuderi_Drago_2004}; the same statement is valid for the brightness temperature in the optically thick case. In contrast, in the optically thin regime the brightness temperature here is slightly larger than for the Maxwellian plasma with the same $T$; see  Fig.~\ref{FigFF}.

\section{Radiation transfer through a gyrolayer in the non-Maxwellian plasmas}
\label{s_transfer}
As has been noted at the end of \S~\ref{s_GR_Max}, the GR source function $S_{f,s}^\sigma$ does not depend on coordinates (for a constant $T$) in a Maxwellian plasma, which simplifies the theory greatly. In particular, the GR emission intensity from a given gyrolayer is described by Eq.~(\ref{Eq_Inten_GR_layer}) regardless of the actual value of the optical depth $\tau$. This is no longer valid for the non-Maxwellian distributions as their source functions do depend on the coordinates within a gyrolayer. This calls for explicit consideration of the radiation transfer through the gyrolayer.

Generation and propagation of emission in a self-absorbing medium is described by the radiation transfer equation \citep[e.g.,][]{FT_2013}
\begin{equation}\label{eqtr}
\frac{\mathrm{d}\mathcal{J}_f^{\sigma}}{\mathrm{d}z}=j_f^{\sigma}-\varkappa^{\sigma}\mathcal{J}_f^{\sigma},
\end{equation}
where we neglect refraction and scattering and assume that the emission modes propagate independently.

GR emissivity and absorption coefficient strongly increase at a gyrolayer where $f\simeq sf_{\mathrm{Be}}$. Therefore, in an inhomogeneous magnetic field,  emission and absorption of radiation  occur primarily within such GR layers. We assume that a GR layer is  narrow (in practice, this implies a somewhat low plasma temperature, $T\lesssim 10^7$~K, see \S~\ref{S_applic} for more detail) so the magnetic field profile along the line of sight within the layer can be approximated by a linear dependence (cf. Eq.~\ref{Eq_B_z_expan}):
\begin{equation}
\frac{f-sf_{\mathrm{Be}}}{f}=\frac{z}{L_{\mathrm{B}}},
\end{equation}
the adjacent GR layers do not overlap, and all other source parameters (except  the magnetic field) are approximately constant within the GR layer.
In this case, for the Maxwellian distribution, the intensity of emission {\it after} passage the GR layer is given by
\begin{equation}\label{eqtrsolve}
\mathcal{J}^{\mathrm{M}, \sigma, \mathrm{out}}_{f, s}=\mathcal{J}^{\mathrm{M}, \sigma, \mathrm{in}}_{f, s}\exp\left(-\tau^{\mathrm{M}, \sigma}_s\right)+S^{\mathrm{M}, \sigma}_{f, s}\left[1-\exp\left(-\tau^{\mathrm{M}, \sigma}_s\right)\right],
\end{equation}
where $\mathcal{J}^{\mathrm{M}, \sigma, \mathrm{in}}_{f, s}$ is the intensity of  emission incident on the gyrolayer {\it from below}  and $S^{\mathrm{M}, \sigma}_{f, s}$ is the source function described by Eq.~(\ref{Eq_Kirchh_Max_Jtau}).

For non-Maxwellian distributions, the radiation transfer equation (\ref{eqtr}) cannot analytically be solved even for the narrow layer approximation adopted, because the corresponding source function vary in space at the GR layers. However, we can write its solution in a form similar to (\ref{eqtrsolve}), namely,
\begin{equation}\label{eqtrnm}
\mathcal{J}^{\sigma, \mathrm{out}}_{f, s}=\mathcal{J}^{\sigma, \mathrm{in}}_{f, s}\exp\left(-\tau^{\sigma}_s\right)+\mathcal{R}_s^{\sigma}
\overline{S^{\sigma}_{f, s}}\left[1-\exp\left(-\tau^{\sigma}_s\right)\right],
\end{equation}
where $\overline{S^{\sigma}_{f, s}}=J^{\sigma}_{f, s}/\tau^{\sigma}_s$ is the effective source function (described by Eq.~(\ref{Eq_Kirchh_n_GR}) for $n$-distribution and Eq.~(\ref{Eq_Kirchh_kappaZ_GR}) for kappa-distribution) and the factor $\mathcal{R}_s^{\sigma}$ is introduced to describe the deviation from the Kirchhoff law. The advantage of this solution form is that the $\mathcal{R}_s^{\sigma}$-factor can be computed once and then used together with the adopted form of the radiation transfer solution, Eq.~(\ref{eqtrnm}). Evidently, this factor  approaches unity in the optically thin limit and when the distribution function approaches the Maxwellian one. In general, the factor $\mathcal{R}_s^{\sigma}$ has to be found numerically. One can note that the first term in Eq. (\ref{eqtrnm}) (describing the GR absorption of the emission produced in the deeper regions) is exactly the same as in Eq. (\ref{eqtrsolve}) because the absorption of the incident radiation is only determined by the total optical depth of the gyrolayer.

By substituting GR emissivity and absorption coefficient (\ref{Eq_emis_kappa}--\ref{Eq_abso_kappa}) for the kappa-distribution into  radiation transfer equation (\ref{eqtr}), introducing a new dimensionless integration variable $t=\zeta_s/\sqrt{2\kappa-3}\propto z$ and having in mind that the solution of the resulting equation should have  general form (\ref{eqtrnm}) at $t\to\infty$ (or $z\to\infty$), we can write the factor $\mathcal{R}_s^{\kappa,\sigma}$ for the kappa-distribution as
\begin{equation}\label{Rf}
\mathcal{R}^{\kappa, \sigma}_s(\tau^{\kappa, \sigma}_s, \kappa-s)=\frac{\tau^{\kappa, \sigma}_s}{1-\exp(-\tau^{\kappa, \sigma}_s)}\frac{u_{\infty}(\tau^{\kappa, \sigma}_s, \kappa-s)}{\sqrt{\pi}}\frac{\Gamma(\kappa-s)}{\Gamma(\kappa-s-1/2)},
\end{equation}
where $u_{\infty}$ is a solution (at $t\to\infty$) of the differential equation
\begin{equation}\label{de}
\frac{\mathrm{d}u(t)}{\mathrm{d}t}=\frac{1}{(1+t^2)^{\kappa-s}}-\frac{\alpha}{(1+t^2)^{\kappa-s+1}}u(t),
\end{equation}
\begin{equation}
\alpha=\frac{\tau^{\kappa, \sigma}_s}{\sqrt{\pi}}\frac{\Gamma(\kappa-s+1)}{\Gamma(\kappa-s+1/2)}
\end{equation}
with the initial condition $u(-\infty)=0$. Note that this factor depends on two parameters only, since the index $\kappa$ and the harmonic number $s$ enter the corresponding expressions in a combination of $\kappa-s$, but not separately. Equation (\ref{de}) has a finite solution at $\kappa-s>1/2$.

\begin{figure*}
\includegraphics{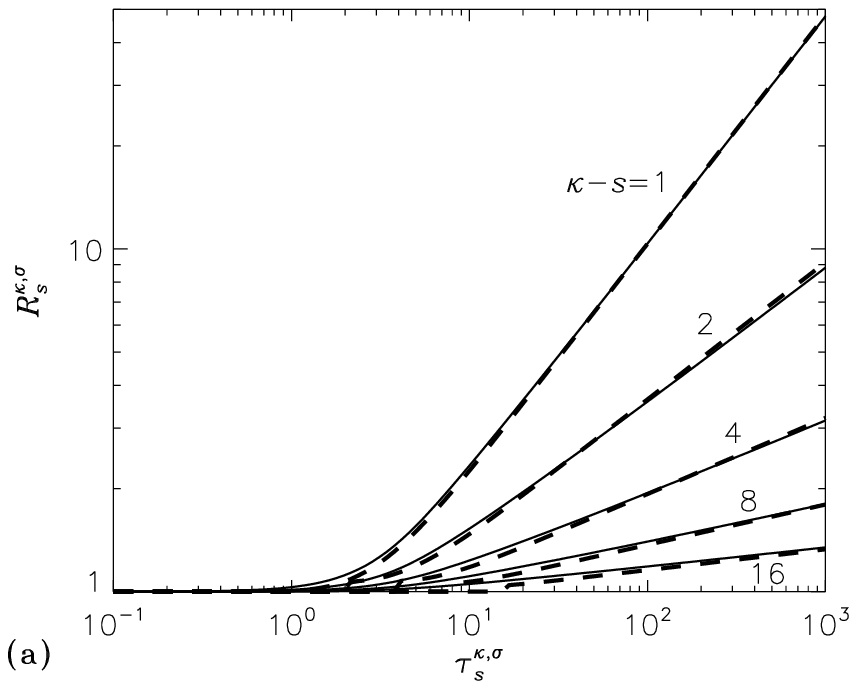}%
\includegraphics{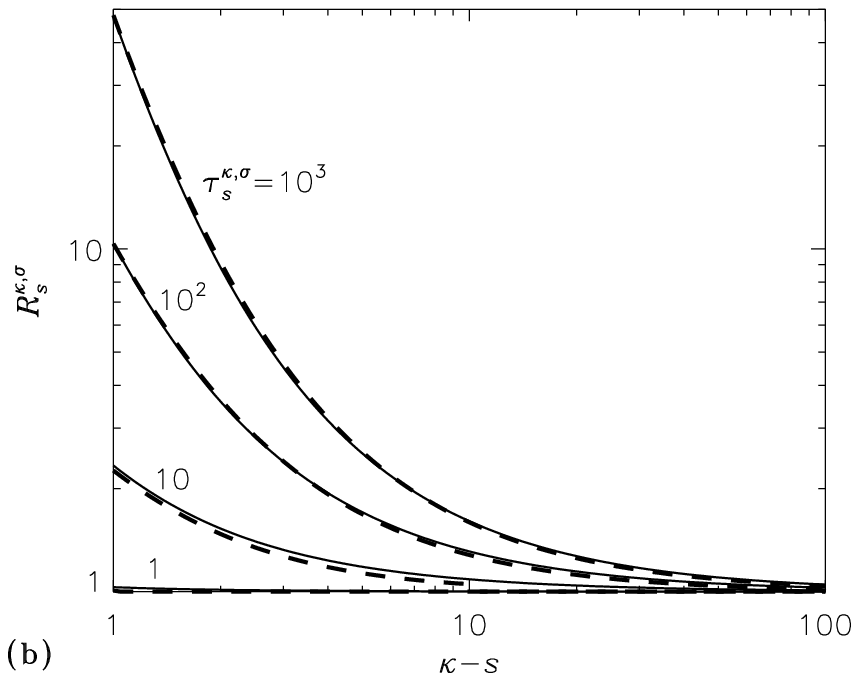}
\caption{Dependence of the correction factor for the kappa-distribution $\mathcal{R}^{\kappa, \sigma}_s(\tau^{\kappa, \sigma}_s, \kappa-s)$ on its parameters. Solid lines: exact values given by Eq. (\protect\ref{Rf}); dashed lines: asymptotical approximation given by Eq. (\protect\ref{eq_R_kappa_analyt}). a) $\mathcal{R}^{\kappa, \sigma}_s$ vs. the optical depth $\tau^{\kappa, \sigma}_s$ for different values of $\kappa-s$; b) $\mathcal{R}^{\kappa, \sigma}_s$ vs. the difference $\kappa-s$ for different values of $\tau^{\kappa, \sigma}_s$.}
\label{FigRKappa}
\end{figure*}

Behavior of $\mathcal{R}^{\kappa, \sigma}_s$-factor as a function of $\tau^{\kappa, \sigma}_s$ and $\kappa-s$ as computed numerically is given in Figs.~\ref{FigRKappa}a,b by solid curves. It is easy to show that the asymptotes of this factor are $\mathcal{R}^{\kappa, \sigma}_s\approx 1$ for $\tau^{\kappa, \sigma}_s<1$ and $\displaystyle\mathcal{R}^{\kappa, \sigma}_s\approx \left[\frac{\tau^{\kappa, \sigma}_s}{3(\kappa-s)^{0.4}}\right]^{\frac{1}{\kappa-s+0.5}}$ for $\tau^{\kappa, \sigma}_s\gg1$. With the use of these two asymptotes one can construct an analytical formula, which correctly describes the $\mathcal{R}^{\kappa, \sigma}_s$-factor in the entire range of interest. A quantitatively accurate approximation is
\begin{equation}
\label{eq_R_kappa_analyt}
\mathcal{R}^{\kappa, \sigma}_s\approx\left\{\begin{array}{l}
1, \qquad \mbox{for} \quad \tau^{\kappa, \sigma}_s<  \kappa-s,\\[6pt]
\displaystyle\left[\frac{\tau^{\kappa, \sigma}_s}{3(\kappa-s)^{0.4}}\right]^{\frac{1}{\kappa-s+0.5}} + \frac{68}{4 + (4+\tau^{\kappa, \sigma}_s)^3} , \qquad \mbox{for} \quad \tau^{\kappa, \sigma}_s>  \kappa-s;
\end{array}
\right.
\end{equation}
the corresponding curves are given by dashed lines in the same figures, which explicitly confirm validity of the approximation.  For the parameter range of $0<\tau^{\kappa, \sigma}_s<10^5$ and $1<\kappa-s<100$ (which covers most cases of interest for solar radio astronomy), a relative error of the analytical approximation (\ref{eq_R_kappa_analyt}) does not exceed 8\%. Thus, with the described modification including the analytical form of the $\mathcal{R}^{\kappa, \sigma}_s$-factor, the gyroresonant theory from the kappa-distribution turns to become almost as simple as that for the Maxwellian distribution.

\begin{figure}
\includegraphics{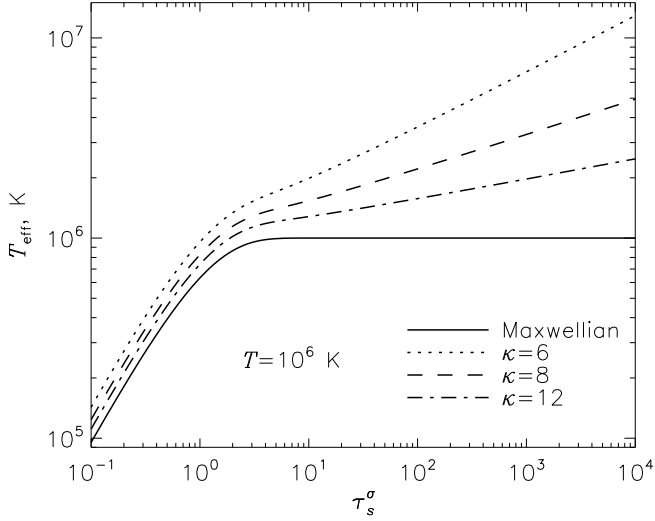}
\caption{Dependence of the brightness temperature of the gyroresonance emission on the optical depth of the gyrolayer for the Maxwellian distribution ($\kappa\to\infty$) and kappa-distributions with different $\kappa$. The plasma temperature is $T=10^6$ K, the cyclotron harmonic number is $s=3$ and the refraction index is $n_{\sigma}\to 1$.}
\label{FigTeffKappa}
\end{figure}

The presence of the non-unitary $\mathcal{R}^{\kappa, \sigma}_s$-factor implies that the brightness temperature of the GR emission from a gyrolayer will depend now on the total optical depth of the gyrolayer. Figure~\ref{FigTeffKappa} displays this dependence for the kappa-distribution with different indices. In contrast with the Maxwellian plasma, for which the brightness temperature is just equal to the plasma kinetic temperature for $\tau\gg 1$, the \textit{\textbf{brightness temperature of the GR emission from a kappa plasma continues to grow with the optical depth $\tau$}}. Not surprisingly, this growth is more pronounced for smaller kappa-indices, i.e., for stronger departure of the plasma from the Maxwellian distribution. The brightness temperature can exceed the kinetic temperature of the kappa plasma by an order of magnitude or even more for a realistic set of parameters.

\begin{figure*}
\includegraphics{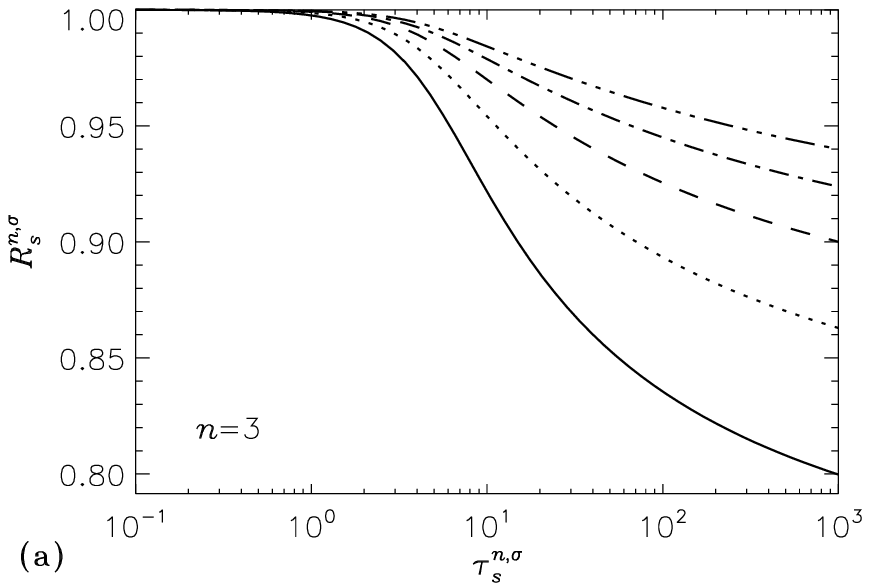}%
\includegraphics{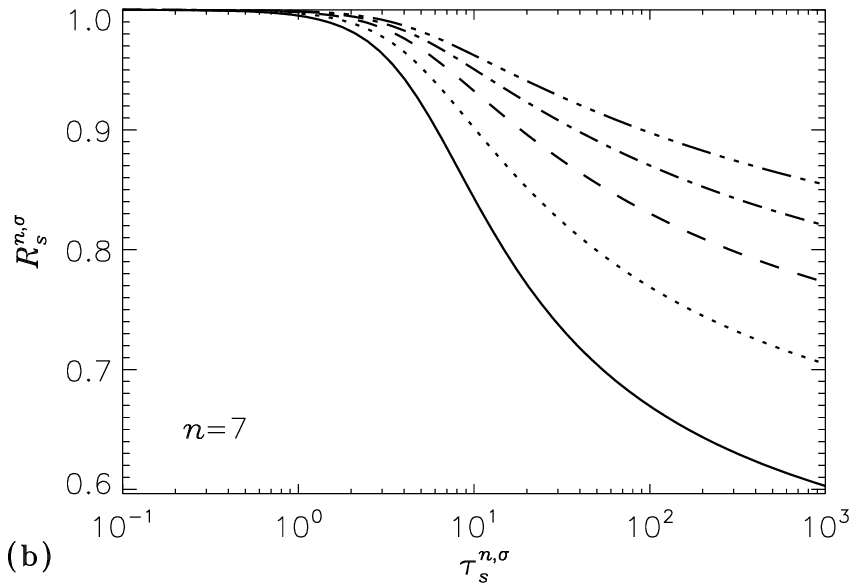}\\
\includegraphics{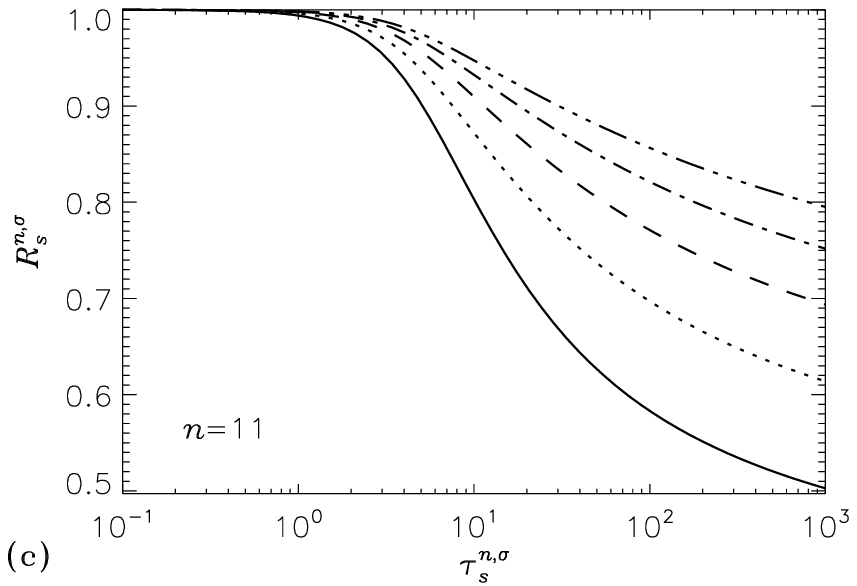}%
\includegraphics{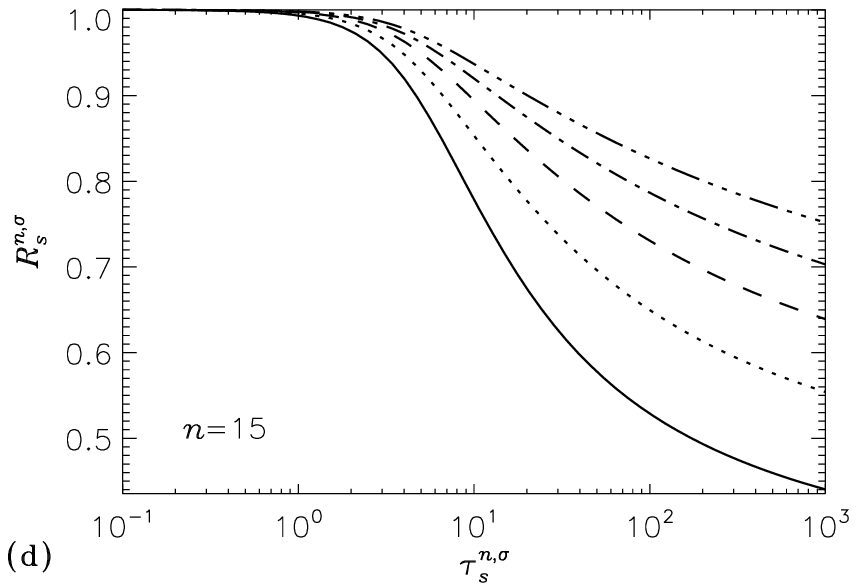}
\caption{Dependence of the correction factor for the $n$-distribution $\mathcal{R}^{n, \sigma}_s(\tau^{n, \sigma}_s, n, s)$ (\protect\ref{Rn}) on the optical depth $\tau^{n, \sigma}_s$ for different values of $n$ and $s$. Different line types correspond to different harmonic numbers: $s=2$ (solid), $s=3$ (dotted), $s=4$ (dashed), $s=5$ (dash-dotted) and $s=6$ (dash-triple-dotted).}
\label{FigRn}
\end{figure*}

For the $n$-distribution, the factor $\mathcal{R}_s^{n,\sigma}$ has the form similar to (\ref{Rf}):
\begin{equation}\label{Rn}
\mathcal{R}^{n, \sigma}_s(\tau^{n, \sigma}_s, n, s)=\frac{\tau^{n, \sigma}_s}{1-\exp(-\tau^{n, \sigma}_s)}\frac{u_{\infty}(\tau^{n, \sigma}_s, n, s)}{\sqrt{\pi}}\frac{\Gamma(s+3/2)}{\Gamma(s+1+n/2)},
\end{equation}
but the differential equation for $u_{\infty}$ is more cumbersome:
\begin{equation}\label{den}
\frac{\mathrm{d}u(t)}{\mathrm{d}t}=\left[e^{-t^2}\sum\limits_{q=0}^l\frac{l!}{s!}\frac{(s+l-q)!}{(l-q)!q!}t^{2q}\right]-\left[\alpha e^{-t^2}\sum\limits_{q=0}^l\frac{l!}{(s-1)!}\frac{(s-1+l-q)!}{(l-q)!q!}t^{2q}\right]u(t),
\end{equation}
with
\begin{equation}
\alpha=\frac{\tau^{n, \sigma}_s}{\sqrt{\pi}}\frac{\Gamma(s+1/2)}{\Gamma(s+n/2)}
\end{equation}
and $l=(n-1)/2$. For $l=0$ ($n=1$, the Maxwellian distribution), as expected, we obtain $\mathcal{R}_s^{(1),\sigma}\equiv 1$. As in this case the $\mathcal{R}_s^{n,\sigma}$-factor is a function of three (rather than two) parameters, it is more convenient here to generate a look-up table of its values, rather than introduce an analytical interpolation, which is more difficult to reliably test in the 3D parameter domain. Such table (providing the relative computation error of less than $2\times 10^{-6}$ for $0<\tau<10^5$, $2\le s\le 20$ and $1\le n\le 15$) is included into our numerical code (see below); the values of the factor $\mathcal{R}^{n, \sigma}_s$ for some subset of the mentioned parameter range are presented in Fig. \ref{FigRn}.

\begin{figure}
\includegraphics{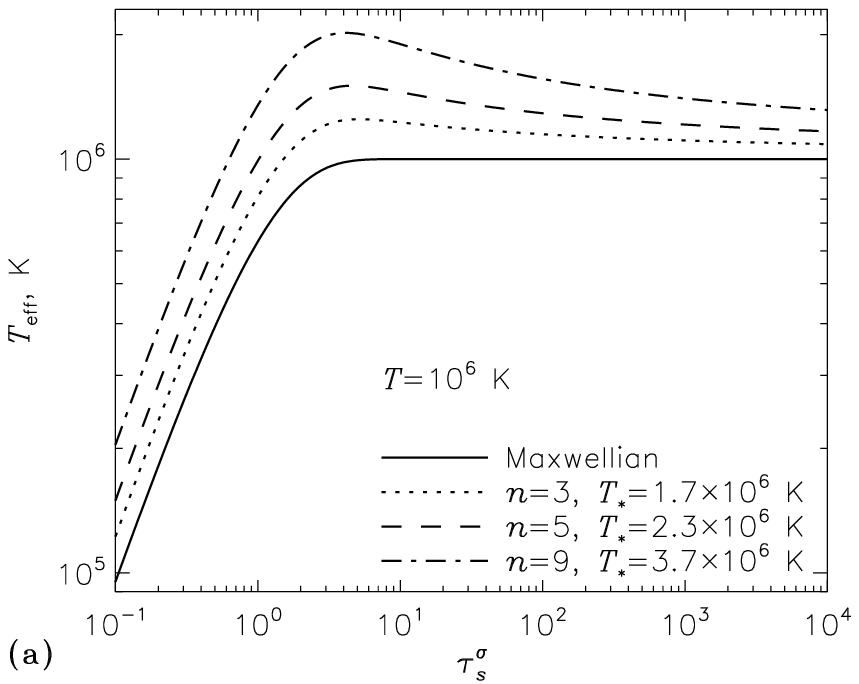} 
\includegraphics{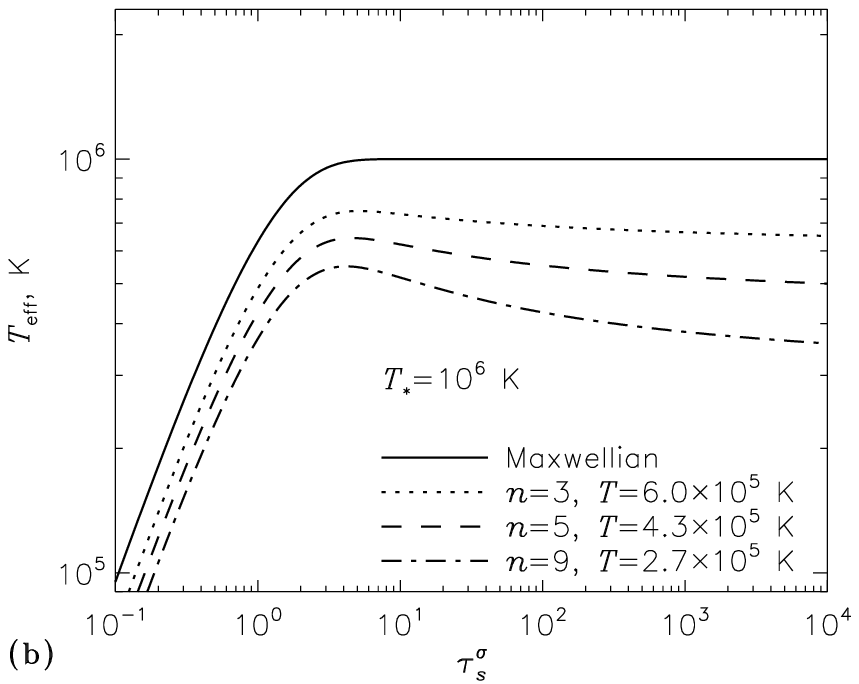}
\caption{Dependence of the brightness temperature of the gyroresonance emission on the optical depth of the gyrolayer for the Maxwellian distribution ($n=1$) and $n$-distributions with different temperatures and $n$-indices. The cyclotron harmonic number is $s=3$ and the refraction index is $n_{\sigma}\to 1$. a) Plots for the constant parameter $T$ ($T=10^6$ K); b) plots for the constant ``pseudo-temperature'' $T_*$ ($T_*=10^6$ K).}
\label{FigTeffN}
\end{figure}

In the case of $n$-distribution the brightness temperature of the GR emission also deviates from the parameter $T$. However, the dependence of $T_{\mathrm{eff}}$ on $\tau$ is nonmonotonic here: $T_{\mathrm{eff}}$ reaches a peak at $\tau\sim 3$ and then starts to decrease. The deviation of $T_{\mathrm{eff}}$ from the $T$ parameter does not exceed a factor of 2-3 for a realistic set of parameters; see Fig.~\ref{FigTeffN}a. For a constant value of $T$, the brightness temperature (both in the optically thick and thin modes) for $n$-distributions is always \textit{higher} than for the Maxwellian one; it increases with increasing $n$. However, this is caused by the  already mentioned fact that the parameter $T$ does not play a role of the effective energy for the $n$-distribution and higher $n$-indices actually correspond to higher average energies of the electrons;  as has been said above, the ``pseudo-temperature'' $T_*=T(n+2)/3$ is a more adequate parameter for comparing the $n$-distributions with different $n$-values and the Maxwellian distribution. As can be seen in Fig.~\ref{FigTeffN}b, for a constant value of $T_*$, the brightness temperature for $n$-distributions is always \textit{lower} than for the Maxwellian one and decreases with increasing $n$.

\section{Application to Active Regions}
\begin{figure}
\includegraphics{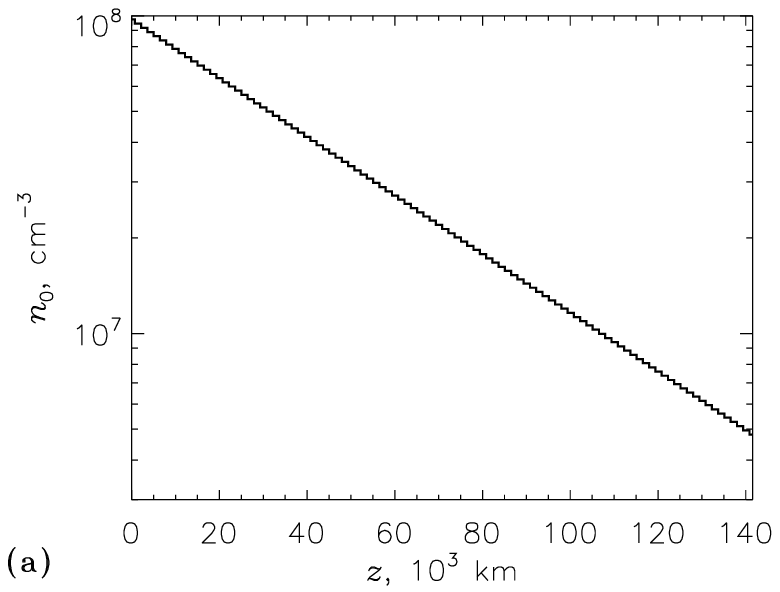} 
\includegraphics{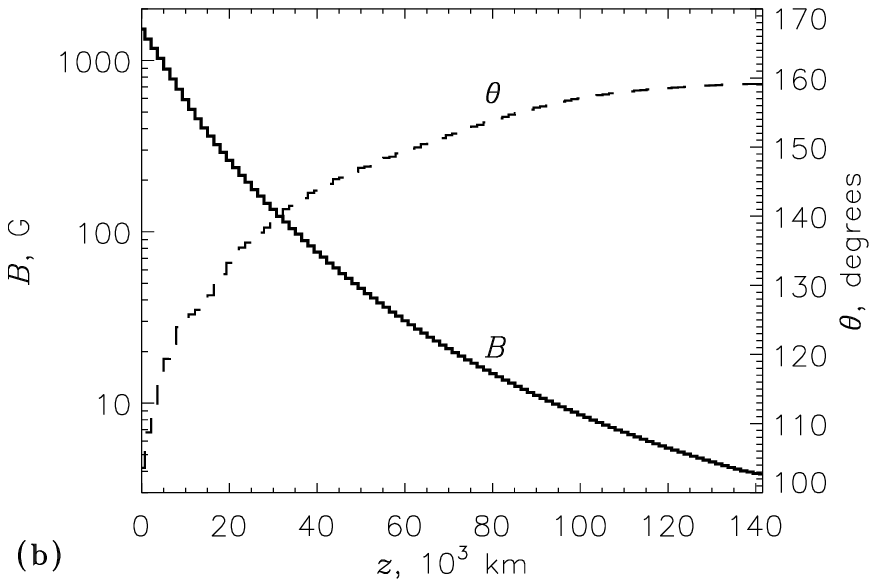}
\caption{Active region model \#1 (simplified): profiles of the plasma density $n_0$, magnetic field strength $B$ and viewing angle $\theta$ along the chosen line of sight. Plasma temperature is $10^6$ K everywhere.}
\label{FigSimProfiles}
\end{figure}

\begin{figure}
\includegraphics{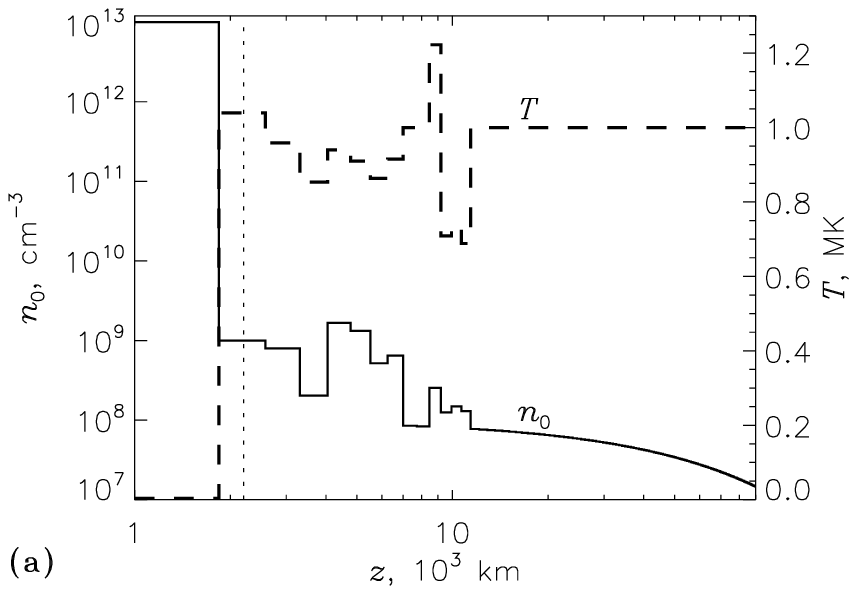} 
\includegraphics{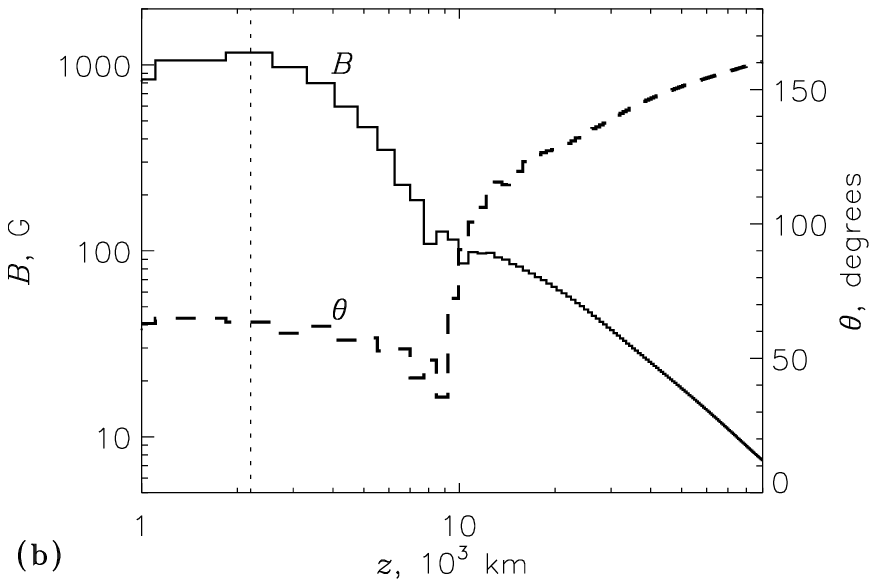}
\caption{Active region model \#2 (advanced): profiles of the plasma density $n_0$, plasma temperature $T$, magnetic field strength $B$ and viewing angle $\theta$ along the chosen line of sight. The abscissa axis is logarithmic to demonstrate better the active region structure at low heights. The vertical dotted lines correspond to the formation layer of the narrowband spectral peaks visible in Figs. \protect\ref{FigRealSpectra}a,c.}
\label{FigRealProfiles}
\end{figure}

Let us consider now how the properties of the radio emission from an active region \citep{Aliss_1984, Akhmedov_etal_1986, Gary_Hurford_1994, Kaltman_etal_1998, Gary_Hurford_2004, gary_keller_2004, Peterova_etal_2006, Lee_2007, RATAN, Tun_etal_2011, Nita_etal_2011, Kaltman_etal_2012} filled with the non-Maxwellian plasmas differ from those in the classical Maxwellian case. To do so we adopt a line-of-sight distribution of all relevant parameters taken from a 3D model we built with our modeling tool, GX Simulator \citep{Nita_GX11b, Nita_GX11a, Nita_GX12}, for a different purpose, and compute the expected emission assuming various energy distribution types of the radiating plasma. At this point we do not address any particular observation but only need a reasonably inhomogeneous distributions of the magnetic field, thermal density, and temperature along the line of sight, implying some complexity of the radio spectrum and polarization. Specifically,  we selected two sets of the line-of-sight distributions of the parameters, which are given in Figs.~\ref{FigSimProfiles} and \ref{FigRealProfiles}.  As can be seen in the figures, all model parameters were defined on a regular grid along the line-of-sight; however, if necessary (e.g., to find the GR layers), a linear interpolation between the grid nodes was used in the simulations.

\begin{figure*}
\begin{center}
\includegraphics{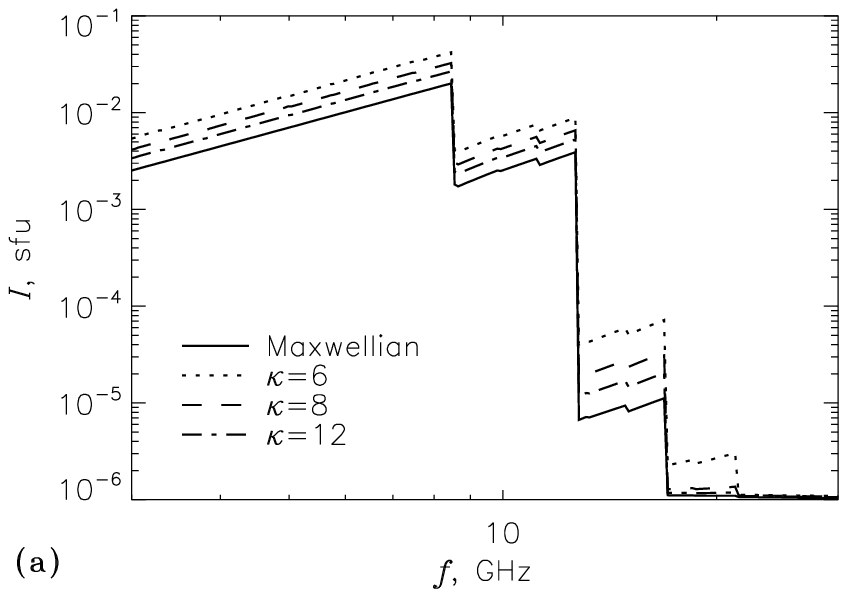}%
\includegraphics{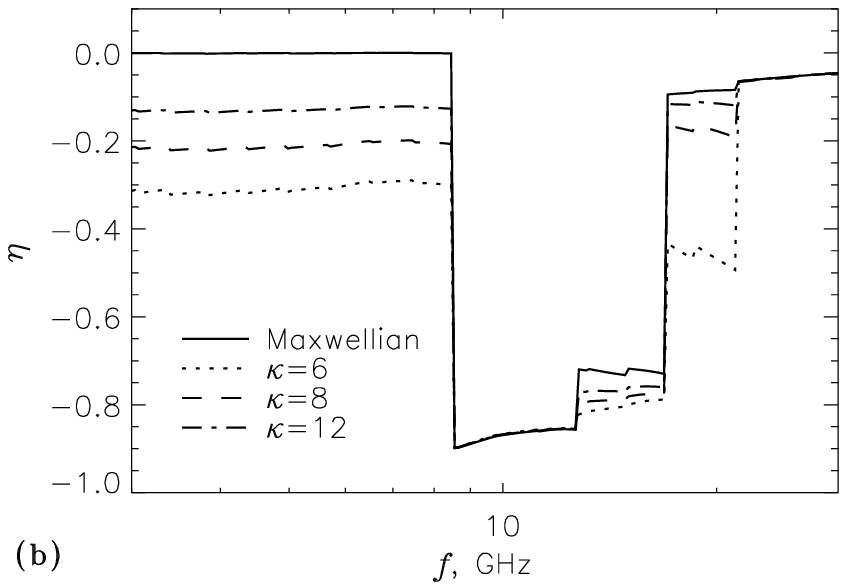}\\
\includegraphics{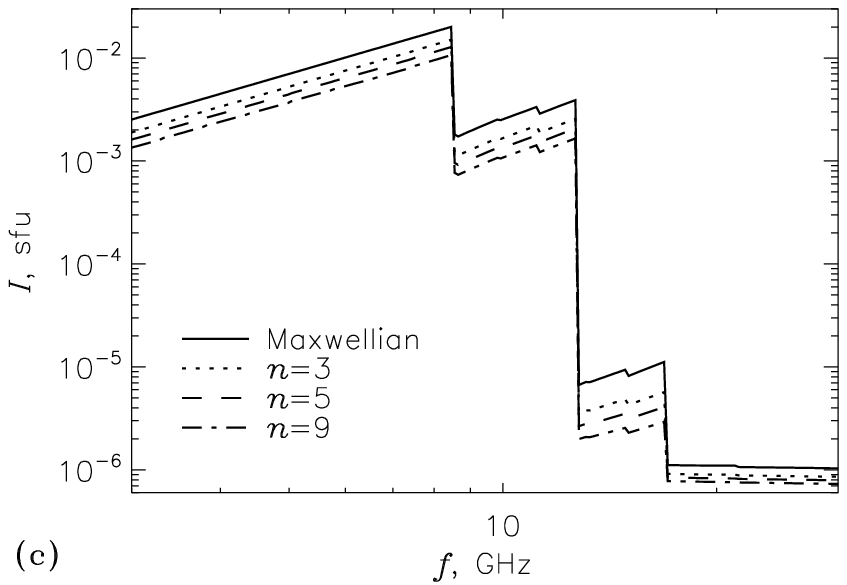}%
\includegraphics{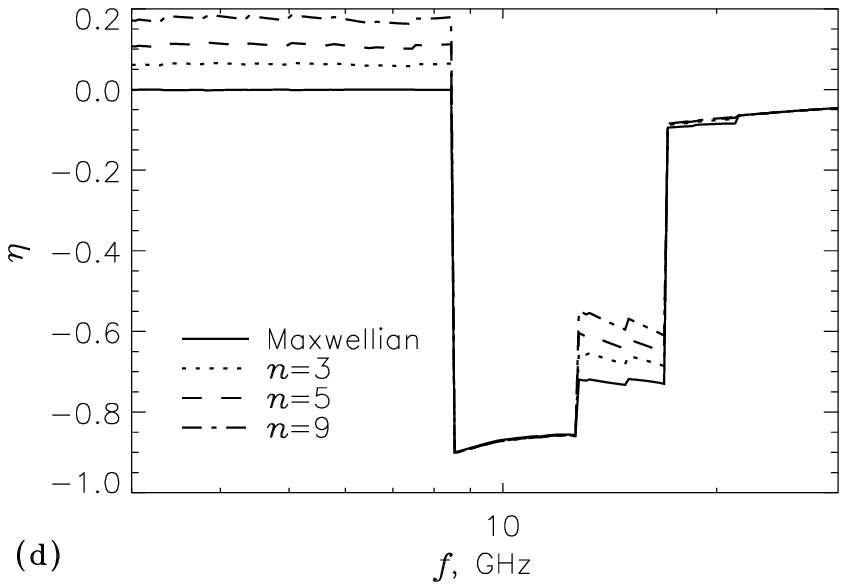}
\end{center}
\caption{Gyroresonance emission spectra (a, c) and polarization (b, d) for the source model shown in Fig. \protect\ref{FigSimProfiles}. Visible source area is 1''$\times$1''. a-b) Maxwellian distribution ($\kappa\to\infty$) and kappa-distributions with different $\kappa$. c-d) Maxwellian distribution ($n=1$) and $n$-distributions with different $n$.}
\label{FigSimSpectra}
\end{figure*}

\begin{figure*}
\begin{center}
\includegraphics{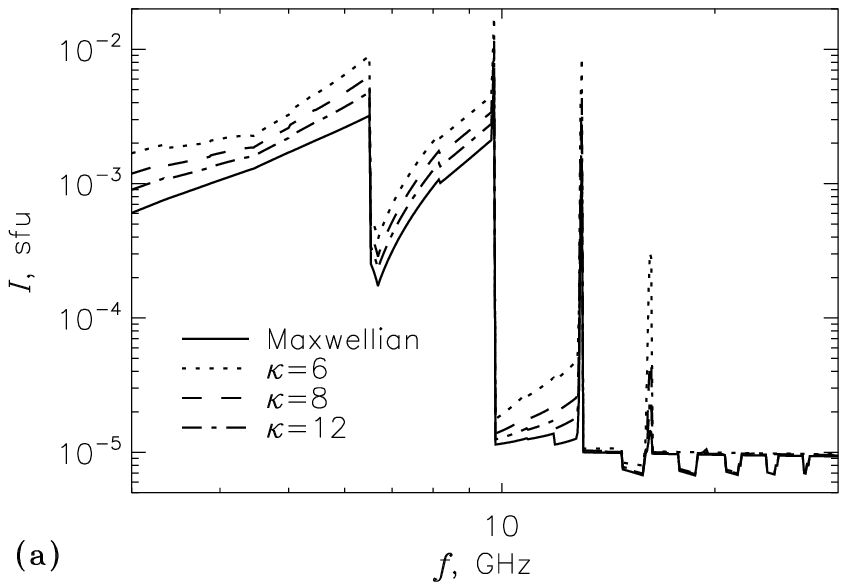}%
\includegraphics{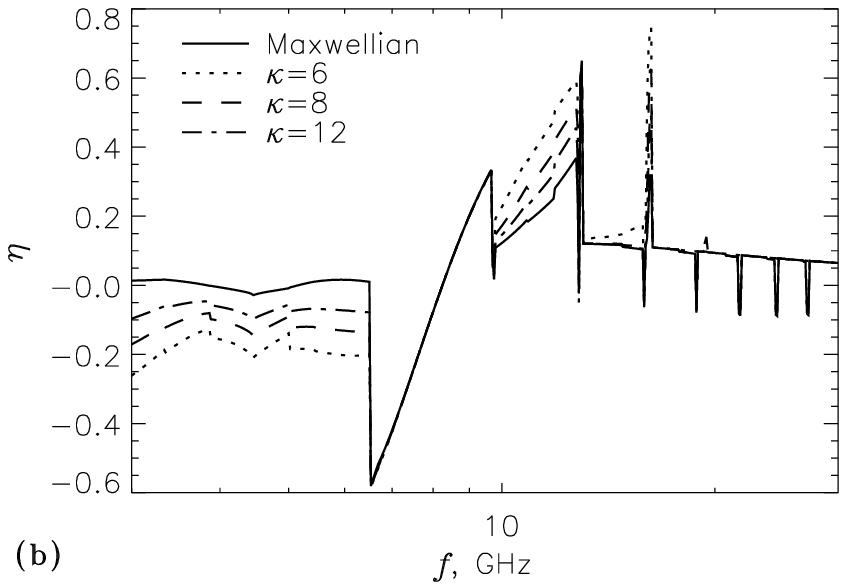}\\
\includegraphics{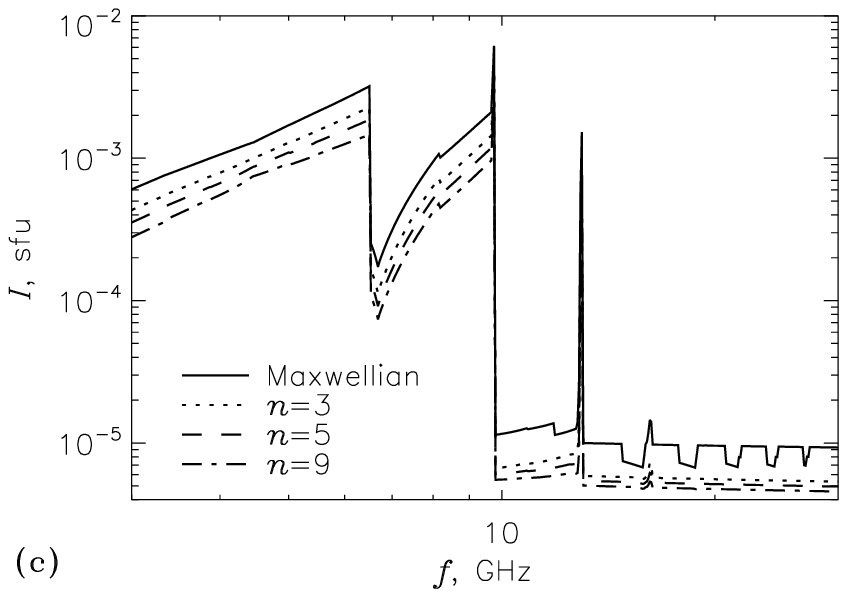}%
\includegraphics{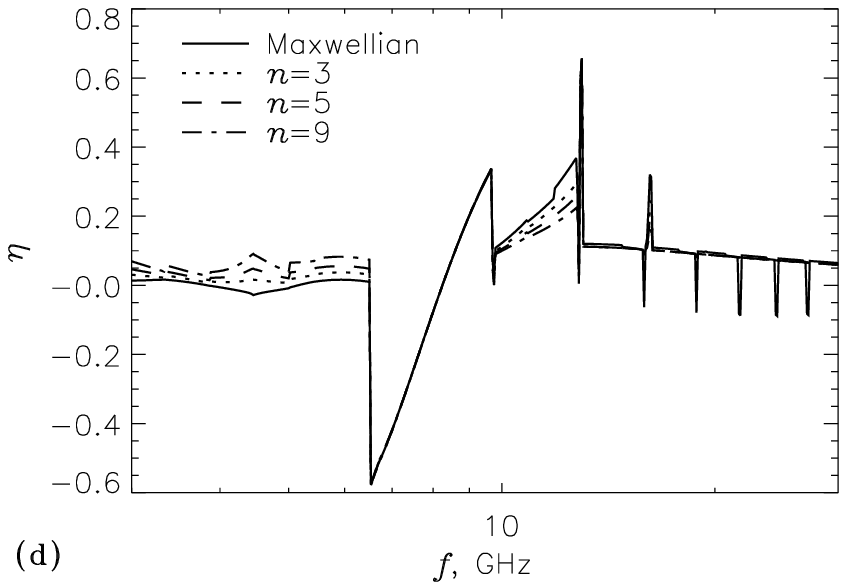}
\end{center}
\caption{Gyroresonance emission spectra (a, c) and polarization (b, d) for the source model shown in Fig. \protect\ref{FigRealProfiles}. Visible source area is 1''$\times$1''. a-b) Maxwellian distribution ($\kappa\to\infty$) and kappa-distributions with different $\kappa$. c-d) Maxwellian distribution ($n=1$) and $n$-distributions with different $n$.}
\label{FigRealSpectra}
\end{figure*}

The first example given in Fig.~\ref{FigSimProfiles} includes a realistic distribution of the magnetic field obtained from an extrapolation of the photospheric magnetic field data, but a simplified hydrostatic distribution of the thermal plasma with a single temperature $T=1$~MK. Figure~\ref{FigSimSpectra} displays the radiation spectra and polarization for the Maxwellian, kappa-, and $n$-distribution with various indices  (for the $n$-distributions, the temperatures corresponding to the constant ``pseudo-temperature'' of $T_*=1$ MK were used). Both GR and free-free processes are included. Not surprisingly, the intensity of the GR emission increases as the kappa-index decreases, which is simply an indication of stronger contribution from the more numerous high-energy electrons from the tail of the kappa-distribution with smaller indices. However, the shape of the spectrum from the kappa-distribution with a given $T$ is difficult to distinguish from that from a Maxwellian plasma with somewhat higher $T$. In contrast, the intensity of the GR emission from the $n$-distributions decreases with increasing $n$-index, but, again, the spectrum shape remains almost the same.

Remarkably, the polarization behavior is distinctly different for the cases of the Maxwellian, kappa- and $n$-distributions. Indeed, at the frequencies where the GR emission is optically thick (below 8 GHz in our example) the degree of polarization from the Maxwellian plasma is zero, see Fig.~\ref{FigSimSpectra}b. This follows from the well known fact that the brightness temperature of the optically thick emission produced by thermal plasma is equal to the kinetic temperature of the plasma in both ordinary and extraordinary wave-modes, which results in a non-polarized emission. The situation is distinctly different for the plasma with the kappa-distribution. As has been shown in \S~\ref{s_transfer}, the brightness temperature of the GR emission from a kappa plasma increases as the optical depth of the gyrolayer increases. It is easy to see that for a given gyrolayer the optical depth of the ordinary mode emission is noticeably smaller than that of the extraordinary mode; thus, the brightness temperature of the extraordinary mode emission is stronger than that of the ordinary mode, which results in a noticeably polarized emission in the sense of extraordinary mode as seen from Fig.~\ref{FigSimSpectra}b. This offers quite a sensitive tool of distinguishing GR emission from kappa- or Maxwellian distributions. A similar effect takes place for $n$-distribution, but   with the opposite (ordinary) sense of polarization (Fig.~\ref{FigSimSpectra}d), because the emission intensity in the optically thick regime decreases with the optical depth; the degree of polarization is slightly lower than for kappa-distribution.

A more realistic inhomogeneous distributions of the plasma density and temperature along the line of sight are demonstrated in Fig. \ref{FigRealProfiles}. In this case, the chromospheric part of the active region (with the plasma density of up to $10^{13}$ $\textrm{cm}^{-3}$ and temperature of 3500 K) is included; the coronal part of the active region contains a number of narrow flux tubes, filled with the thermal plasma according to a nanoflare heating model \citep{Klimchuk_etal_2008, Klimchuk_etal_2010}, which makes the height profiles of all parameters non-monotonic. In addition, the sign of the projection of the magnetic field vector on the line-of-sight experiences a reversal within the active region. As one might expect, the radiation spectra and polarization (see Fig. \ref{FigRealSpectra}) become now more diverse and structured. There is a polarization reversal at the frequency of about 8 GHz, caused by the frequency-dependent mode coupling  at the layer with the transverse magnetic field (roughly, at the level of 10\,000 km above the photosphere, see Fig.~\ref{FigRealProfiles}b). Furthermore, there are several sharp narrowband peaks at the harmonically-related frequencies in the intensity spectra. These peaks are produced at the bottom of the corona, at the layer where  magnetic field reaches its maximum along the line of sight ($B\simeq 1160$~G, so that the peaks at 9.7, 13.0, and 16.2 GHz correspond to the third, fourth, and fifth gyroharmonic, respectively), the plasma density is relatively high ($n_0\simeq 10^9$ $\textrm{cm}^{-3}$), while the plasma temperature ($T\simeq 1$ MK) is typical of the corona. This parameter combination is indicative that our line of sight crosses one of the closed flux tubes located in this instance at the base of corona.
Strong magnetic field, high plasma density and temperature and small local gradient of the magnetic field render the emission to be optically thick at the third and partially fourth harmonic with the brightness temperature about the plasma kinetic temperature. For the same reason, emission intensity at even fifth harmonic is relatively large. Narrow height extension of this region results, however, in quite a narrowband emission. A more broadband feature can be seen around 8~GHz, which corresponds to a few voxels with a dense plasma and the field around 800~G. Note that there is at least one hotter flux tube higher in the corona with $T\simeq1.2$~MK, but these tubes have no visible effect on the spectra in Fig. \ref{FigRealSpectra} because their magnetic fields are too weak ($\sim 100$~G) to produce the gyroemission above 1 GHz; however, they could be distinguished at lower frequencies. The mentioned fine spectral structures can be potentially used as a very precise tool for measuring magnetic fields in such fluxtubes; however, this requires observations with high spectral and angular resolutions (since the peaks in the spatially integrated spectra can be smoothed due to the source inhomogeneity \textit{across} the line-of-sight). Like in the previous model, the intensity spectra for different electron distributions have similar shapes, although in some frequency ranges (e.g., at $f\gtrsim 10$ GHz in Fig. \ref{FigRealSpectra}a) they can demonstrate noticeably different slopes; the polarization remains much more sensitive to the electron distribution type and parameters than the radiation intensity.

\section{Applicability of the GR approximation}
\label{S_applic}
\begin{figure}
\includegraphics{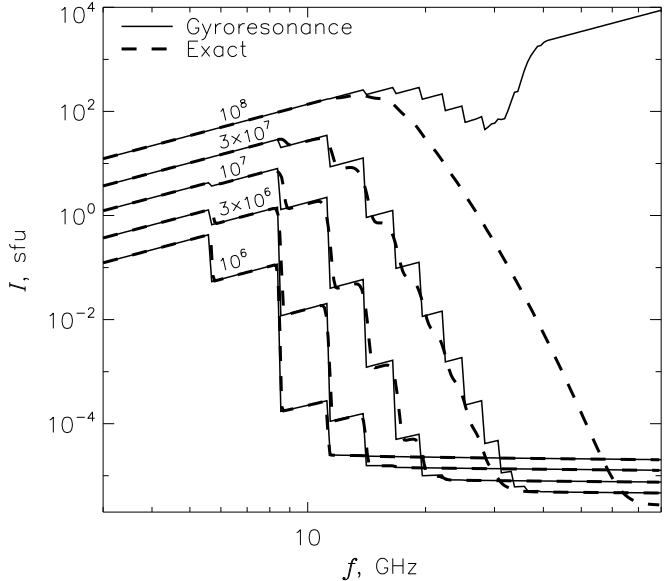}
\caption{Simulated emission spectra for the Maxwellian distributions with different temperatures. Solid lines: results obtained using the gyroresonance approach described in this paper; dashed lines: results obtained using precise relativistic gyrosynchrotron formulae. The temperatures (in Kelvins) are indicated by numbers near the lines; other source parameters are given in the text.}
\label{FigLMaxwell}
\end{figure}

\begin{figure*}
\begin{center}
\includegraphics{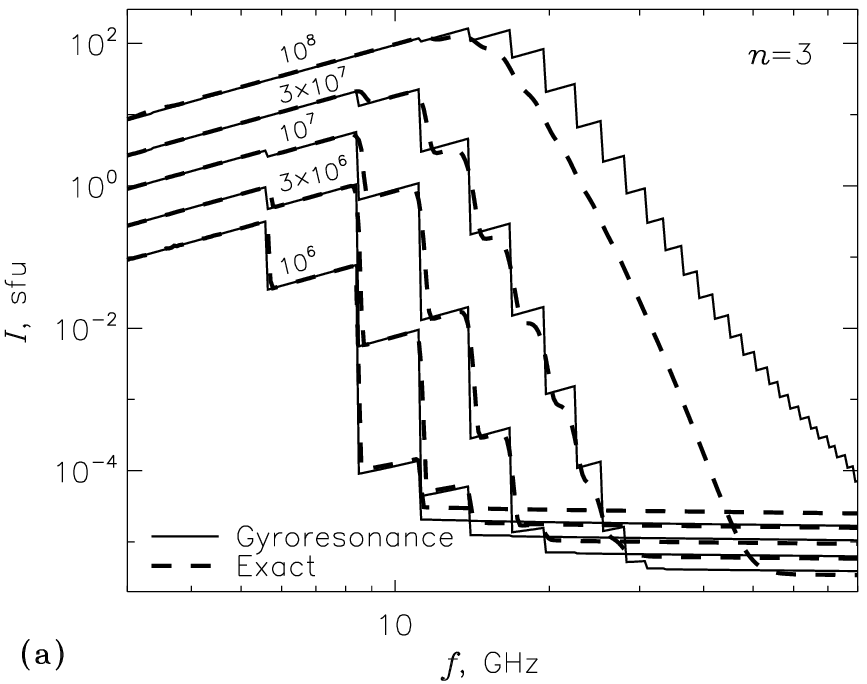}%
\includegraphics{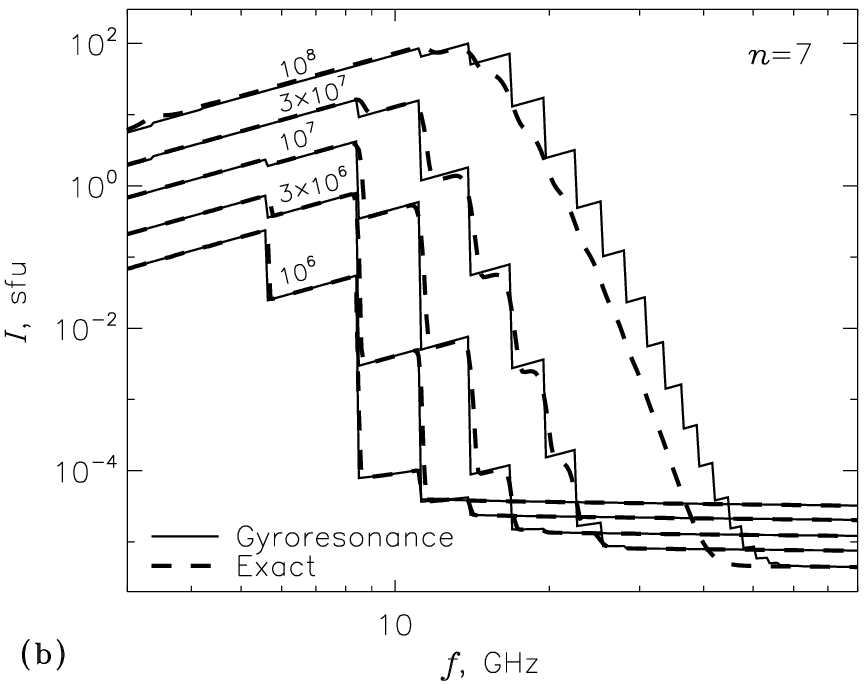}
\end{center}
\caption{Same as in Fig. \protect\ref{FigLMaxwell}, for the $n$-distributions with different temperatures and $n$-indices. The numbers near the lines indicate the effective ``pseudo-temperatures'' $T_*$.}
\label{FigLN}
\end{figure*}

\begin{figure*}
\begin{center}
\includegraphics{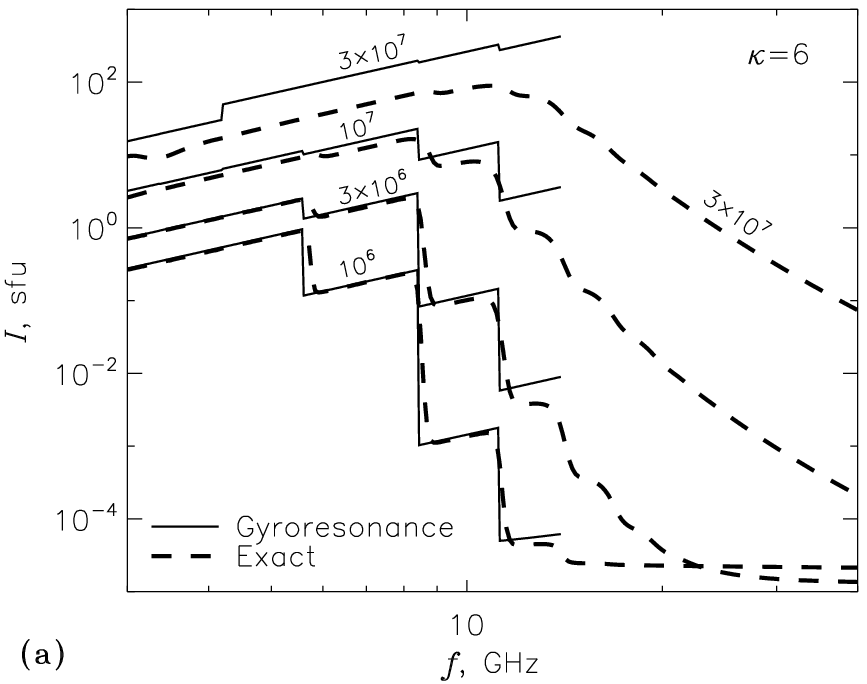}%
\includegraphics{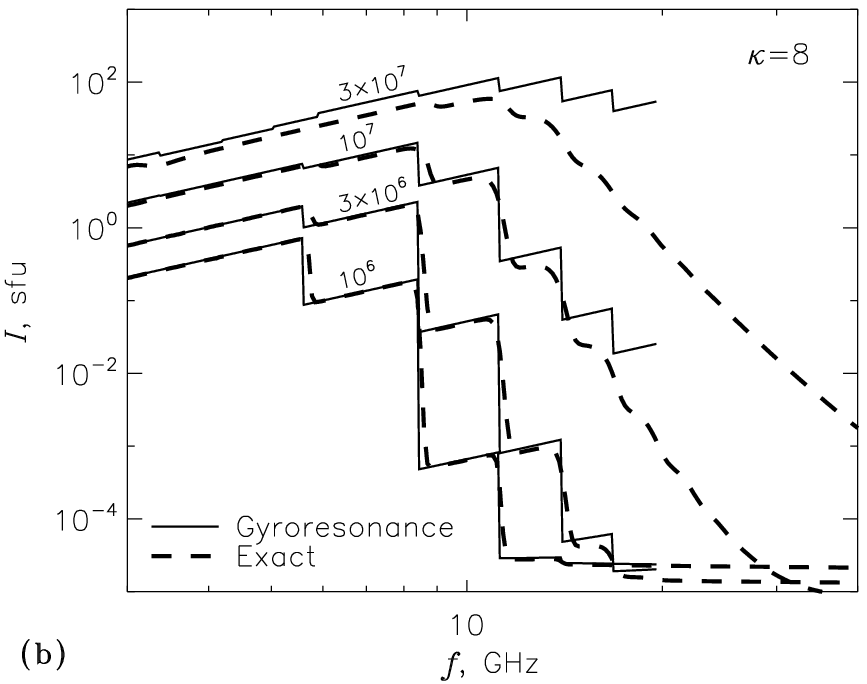}\\
\includegraphics{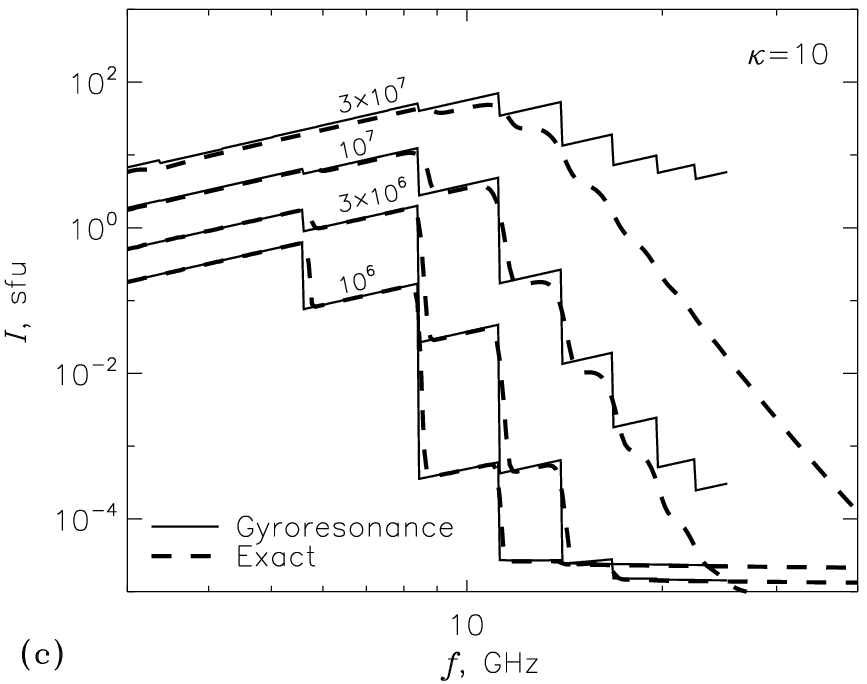}%
\includegraphics{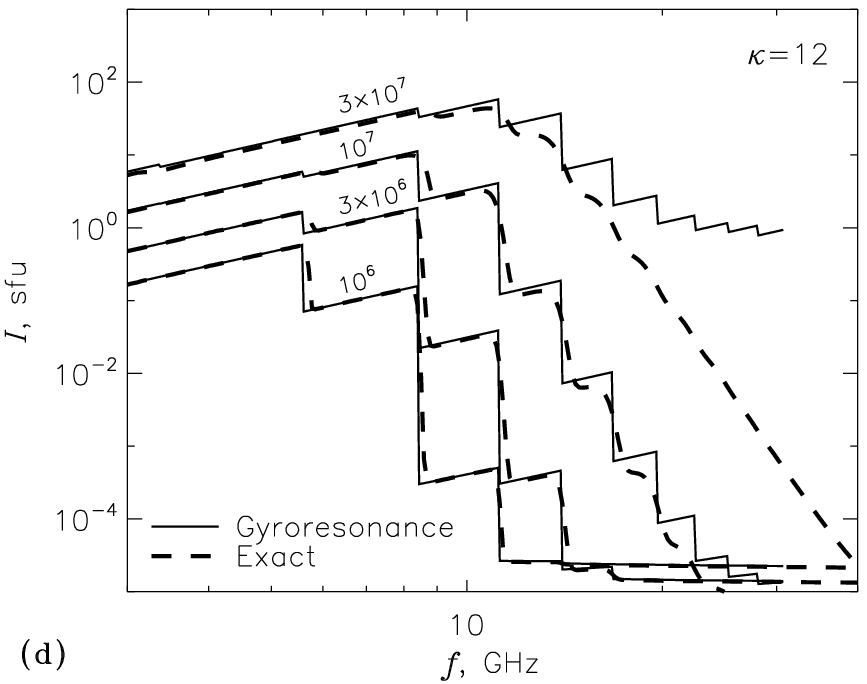}
\end{center}
\caption{Same as in Fig. \protect\ref{FigLMaxwell}, for the kappa-distributions with different temperatures and kappa-indices. The gyroresonance approach (shown by solid lines) is available only at $s<\kappa-1/2$.}
\label{FigLKappa}
\end{figure*}

Let us address now the applicability of the considered here GR approximation.  For this purpose, we have compared the numerical results obtained using the approximate (gyroresonance) and exact (gyrosynchrotron) formulae (see Figs. \ref{FigLMaxwell}--\ref{FigLKappa}). The simulations were performed for a model emission source with homogeneous plasma density and temperature and constant magnetic field direction; in all simulations we used the plasma density of $n_0=10^8$ $\textrm{cm}^{-3}$ and viewing angle of $\theta=60^{\circ}$. The magnetic field strength varied linearly with the distance along the line-of-sight (from 1000 to 300 G over the source depth of 10\,000 km, which corresponds to the inhomogeneity scale of $L_{\mathrm{B}}\simeq 9300$ km). The gyrosynchrotron emission was calculated using the exact relativistic formulae \citep{Eidman_1958, Eidman_1959, mel68, Ramaty_1969} in the form given by \citet{mel68} implemented into the fast codes by \citet{Fl_Kuzn_2010}; the integration step along the line of sight was  manually chosen to be small enough to resolve the gyroresonance levels. 
The emission intensities shown in the figures correspond to the source located at the Sun, with the visible area of $10^8$ $\textrm{km}^2$.

We can see that for the Maxwellian distribution (Fig. \ref{FigLMaxwell}), the GR approximation ensures excellent accuracy up to the plasma temperature of about 10 MK. At higher temperatures ($\gtrsim 30$ MK) the approximation correctly reproduces the optically thick part of the spectrum, but tends to overestimate the optically thin emission intensity; in addition, the exact optically-thin spectra are smoother than the approximate ones. At very high temperatures and/or frequencies (e.g., at $f\gtrsim 30$ GHz for $T=100$ MK), the GR approximation completely breaks down, because the condition $\lambda\ll 1$ (that was used to get rid of the Bessel functions) is no longer valid. The limiting frequency increases with the plasma temperature decrease: for the above mentioned parameters, the GR approximation is valid up to the frequencies of about 120 and 450 GHz for the temperatures of 30 and 10 MK, respectively. Since the gyroemission intensity (for the typical coronal temperatures) at those frequencies is negligible, the GR approximation seems to be well sufficient for solar applications.

Similar conclusions can be drawn for the $n$-distributions (see Fig. \ref{FigLN}). The applicability range of the GR approximation is now even wider than for the Maxwellian distribution, because for the same average energies of the electrons (characterized by the ``pseudo-temperature'' $T_*$) the $n$-distributions have lower $T$ parameters, which implies a steeper decrease of the distribution function at high energies.

The situation is somewhat different for kappa-distribution as it has a power-law tail at high energies; this is why the GR approximation can only be valid up to certain low harmonics determined by the $\kappa$ index, as has already been noted earlier. Figure \ref{FigLKappa} gives a good idea about the applicability region of the GR approximation for the kappa-distribution  where both the temperature and kappa-index are important. In general, the region of the GR approximation applicability is narrower for the kappa- than for the Maxwellian distribution. For very energetic plasmas (e.g., at $\kappa\lesssim 8$ and $T\gtrsim 30$ MK), even the optically thick emission is badly reproduced. As expected, the smaller the kappa-index the smaller the highest temperature at which the GR formulae can be used. Another limitation is the frequency: although the GR approximation breaks formally down only above $f/f_{\mathrm{Be}}\simeq s = \mbox{Integer}[\kappa+1/2]$, it becomes increasingly inaccurate when approaching this boundary from below; this effect is more pronounced for higher temperatures, since the corresponding emission spectra extend to higher frequencies. Overall, the use of the developed here GR theory is safe for the temperatures below 3 MK for the modest values of $\kappa\gtrsim 7$, which can be expected in a steady-state plasma of solar active regions. For even higher kappa-indices, the distribution properties approach those of the Maxwellian distribution, and for $\kappa>12$ the GR approximation becomes valid up to about 10 MK.

\section{Discussion} 
This paper has developed analytical theory of the GR and free-free emission from plasmas characterized by non-Maxwellian isotropic kappa- or $n$-distributions. In particular, we demonstrated that the free-free emission from $n$-distribution is always optically thin, while the brightness temperature of the optically thick free-free emission from a plasma with kappa-distribution is lower than that for the Maxwellian plasma with the same $T$. We emphasize that the use of these new formulae is needed any time when an emission from such non-Maxwellian plasma is modelled. For example, if one considers the gyrosynchrotron or plasma emission from a plasma with kappa-distribution, then the free-free emission and absorption must although be computed for the same kappa-distribution; the use of the standard free-free formulae can lead to inconsistent results.

{We note that for some solar flares the coronal X-ray emission can equally well be fitted by either kappa- or a Maxwellian core plus a power-law distribution \citep{Kasparova_Karlicky_2009, Oka_etal_2013}, which can also be presented in the form of thermal-nonthermal (TNT) distribution \citep{Holman_Benka_1992, Benka_Holman_1994}. This implies that the results obtained here for kappa distribution can, to some extent, apply to the TNT distribution. However, this analogy is limited for the following two reasons. First, the TNT distribution is only similar to the kappa distribution if the power-law tail of the TNT distribution contains a significant fraction of the total number of the electrons at the source. And second, during a flare the plasma temperature is often high, larger than 10 MK, and the nonthermal tail spectra are relatively hard, so the GR approximation breaks down (see \S~\ref{S_applic}), so a more general gyrosynchrotron treatment \citep{Fl_Kuzn_2010} has to be used.}

The GR emission from the non-Maxwellian plasmas is distinctly different from the classical Maxwellian one. These {differing properties} can be used to observationally distinguish the active region plasmas with either Maxwellian or non-Maxwellian distributions---in particular, by using the polarization data, which are shown to be strongly different in the case of kappa-distribution compared with the Maxwellian one. We emphasize that the microwave imaging spectroscopy and polarimetry measurements are supposed to be extraordinary sensitive to the distribution type, so the validity of (or deviation from) the Maxwellian distribution of the coronal plasma in the active regions can in principle be tested with a very high accuracy.

The theory presented here has been implemented into an efficient computer code, which we compiled, in particular, as a dynamic link library  (Windows DLL) callable from IDL. This  library is included in the  SolarSoft (SSW) distribution of our simulation tool, GX Simulator, which is publicly available. The library, along with sample calling routines, is subject of data sharing and so can be obtained upon request.

\acknowledgments
This work was supported in part by NSF grants AST-0908344 and
AGS-1250374 and NASA grants  NNX11AB49G and NNX13AE41G to New
Jersey Institute of Technology, by the
Russian Foundation of Basic Research (grants 12-02-00173, 12-02-00616, 12-02-91161, 13-02-10009 and 13-02-90472) and by a Marie Curie International Research Staff Exchange Scheme Fellowship within the 7th European Community Framework Programme. This work also benefited from workshop support from the
International Space Science Institute (ISSI).

\bibliographystyle{apj}
\bibliography{AR,AR_bib,fleishman,Ch_Drago_2004}
\end{document}